%% file: Main_v9.tex
\documentclass[a4paper,12pt]{article} 
\pdfoutput=1               
\usepackage{jcappub_MG}    

\usepackage{amsmath}
\usepackage{amsfonts}
\usepackage{amsxtra}
\usepackage{hyperref}
\usepackage{amssymb}
\usepackage{dsfont}
\usepackage{graphicx,epsfig}
\usepackage{natbib}
\usepackage{xspace}

\newcommand{\lsi}{\,\raisebox{-0.13cm}{$\stackrel{\textstyle<}{\textstyle\sim}$}\,}
\newcommand{\gsi}{\,\raisebox{-0.13cm}{$\stackrel{\textstyle>}{\textstyle\sim}$}\,}
\def\fermi{\textsl{Fermi}\xspace}
\def\veritas{\textsl{VERITAS}\xspace}
\def\magic{\textsl{MAGIC}\xspace}
\def\lat{{LAT}\xspace}
\def\wmap{\textsl{WMAP}\xspace}

\def\hess{\textsl{H.E.S.S.}\xspace}
\def\magic{\textsl{MAGIC}\xspace}

\def\planck{\textsl{Planck}\xspace}
\def\galprop{\texttt{GALPROP}\xspace}
\def\eg{{e.g.}\@\xspace}
\def\ie{{i.e.}\@\xspace}

\newcommand{\ltsima} {$\; \buildrel < \over \sim \;$}
\newcommand{\gtsima} {$\; \buildrel > \over \sim \;$}
\newcommand{\lta} {\lower.5ex\hbox{\ltsima}}
\newcommand{\gta} {\lower.5ex\hbox{\gtsima}}
\def\la{\langle}
\def\ra{\rangle}
\def\sigv{\la \sigma v\ra}
\def\de{$^{\circ}$}
\def\Msun{{\rm M}_\odot}
\def\Mmin{M_{\rm min}}
\def\kmax{k_{\rm max}}
\newcommand{\be}{\begin{equation}}
\newcommand{\ee}{\end{equation}}
\newcommand{\beq}{\begin{equation}}
\newcommand{\eeq}{\end{equation}}
\newcommand{\bea}{\begin{eqnarray}}
\newcommand{\eea}{\end{eqnarray}}
\newcommand{\beaa}{\begin{eqnarray*}}
\newcommand{\eeaa}{\end{eqnarray*}}
\newcommand{\ba}{\begin{array}}
\newcommand{\ea}{\end{array}}
\newcommand{\bi}{\begin{itemize}}
\newcommand{\ei}{\end{itemize}}
\newcommand{\ben}{\begin{enumerate}}
\newcommand{\een}{\end{enumerate}}

\ifx\pdfoutput\undefined
  \usepackage[dvips,bookmarks]{hyperref}	% This is for arXiv.org
\else
  \usepackage{hyperref}	                % This is for pdftex
\fi
\hypersetup{colorlinks,bookmarksopen,bookmarksnumbered,citecolor=rossoc,linkcolor=blue,pdfstartview=FitH,urlcolor=rossos}

\definecolor{rossos}{cmyk}{0,1,1,0.55}
\definecolor{rossoc}{cmyk}{0,1,1,0.2}
\usepackage{color}

\title{Limits on Dark Matter Annihilation Signals from the Fermi LAT 4-year Measurement of the Isotropic Gamma-Ray Background}
\date{\today}

\subheader{ULB/14-21}

\collaboration{The Fermi LAT Collaboration}

\author [1]  {M.~Ackermann}
\author [2]  {M.~Ajello}
\author [3]  {A.~Albert}
\author [4]  {L.~Baldini}
\author [5,6]  {G.~Barbiellini}
\author [7,8]  {D.~Bastieri}
\author [9]  {K.~Bechtol}
\author [4]  {R.~Bellazzini}  
\author [10]  {E.~Bissaldi}
\author [3]  {E.~D.~Bloom}
% ...

\affiliation[1]{Deutsches Elektronen Synchrotron DESY, D-15738 Zeuthen, Germany}
\affiliation[2]{Department of Physics and Astronomy, Clemson University, Kinard Lab of Physics, Clemson, SC 29634-0978, USA}
\affiliation[3]{W. W. Hansen Experimental Physics Laboratory, Kavli Institute for Particle Astrophysics and Cosmology, Department of Physics and SLAC National Accelerator Laboratory, Stanford University, Stanford, CA 94305, USA}
\affiliation[4]{Istituto Nazionale di Fisica Nucleare, Sezione di Pisa, I-56127 Pisa, Italy}
\affiliation[5]{Istituto Nazionale di Fisica Nucleare, Sezione di Trieste, I-34127 Trieste, Italy}
\affiliation[6]{Dipartimento di Fisica, Universit\`a di Trieste, I-34127 Trieste, Italy}
\affiliation[7]{Istituto Nazionale di Fisica Nucleare, Sezione di Padova, I-35131 Padova, Italy}
\affiliation[8]{Dipartimento di Fisica e Astronomia ``G. Galilei'', Universit\`a di Padova, I-35131 Padova, Italy}
\affiliation[9]{Kavli Institute for Cosmological Physics, University of Chicago, Chicago, IL 60637, USA}
\affiliation[10]{Istituto Nazionale di Fisica Nucleare, Sezione di Trieste, and Universit\`a di Trieste, I-34127 Trieste, Italy}
% ...
\emailAdd{afrancko@slac.stanford.edu, michael.gustafsson@theorie.physik.uni-goettingen.de, %michael.gustafsson@ulb.ac.be, 
sanchezconde@fysik.su.se, gabrijela.zaharijas@ung.si}

\abstract{We search for evidence of dark matter (DM) annihilation in the isotropic gamma-ray background (IGRB) measured with 50 months of \fermi Large Area Telescope (LAT) observations. An improved theoretical description of the cosmological DM annihilation signal, based on two complementary techniques and assuming generic weakly interacting massive particle (WIMP) properties, renders more precise predictions compared to previous work. More specifically, we estimate the cosmologically-induced gamma-ray intensity to have an uncertainty of a factor $\sim 20$ in canonical setups. We consistently include both the Galactic and extragalactic signals under the same theoretical framework, and study the impact of the former on the IGRB spectrum derivation. We find no evidence for a DM signal and we set limits on the DM-induced isotropic gamma-ray signal. Our limits are competitive for DM particle masses up to tens of TeV and, indeed, are the strongest limits derived from \fermi~\lat data at TeV energies. This is possible thanks to the new \fermi~LAT IGRB measurement, which now extends up to an energy of 820 GeV. We quantify uncertainties in detail and show the potential this type of search offers for testing the WIMP paradigm with a complementary and truly cosmological probe of DM particle signals.} 

\include{aastex_journal_defs}

\begin{document}

\maketitle

\input{1IntroSHORT}

\input{2Theory}
\input{21HaloModel}

\input{22PowerSpectrum}

\input{23HM_PS_Comparison}

\input{24GalacticSS}

\input{3DMlimits}
\input{4GadgetchecksSHORT}

\input{5Summary}

\acknowledgments
The \fermi LAT Collaboration acknowledges generous ongoing support from a number of agencies and institutes that have supported both the development and the operation of the LAT as well as scientific data analysis. These include the National Aeronautics and Space Administration and the Department of Energy in the United States, the Commissariat l'Energie Atomique and the Centre National de la Recherche Scientifique, Institut National de Physique Nucl\'eaire et de Physique des Particules in France, the Agenzia Spaziale Italiana and the Istituto Nazionale di Fisica Nucleare in Italy, the Ministry of Education, Culture, Sports, Science and Technology (MEXT), High Energy Accelerator Research Organization (KEK) and Japan Aerospace Exploration Agency (JAXA) in Japan, and the K. A. Wallenberg Foundation, the Swedish Research Council and the Swedish National Space Board in Sweden. Additional support for science analysis during the operations phase is gratefully acknowledged from the Istituto Nazionale di Astrofisica in Italy and the Centre National d'Etudes Spatiales in France.

M.G.\ is supported by the Belgian Science Policy (IAP VII/37), the IISN and the ARC project. M.A.S.C. acknowledges the support of the Wenner-Gren foundation to develop his research and the NASA grant NNH09ZDA001N for the study of the extragalactic background. G.Z.\ is grateful to SLAC for hospitality during part of the realization of this work. The authors are thankful to Emiliano Sefusatti for help with producing some of the figures. We also thank Mattia Fornasa for useful discussions and comments.

Some of the results in this paper have been derived using the HEALPix code \cite{Gorski:2004by}.

\appendix
\input{appendixA_limitsABC}
\clearpage

\input{appendixB_3sigma}
\clearpage

%\bibliography{CosmoWIMPpapers}         
\providecommand{\href}[2]{#2}\begingroup\raggedright\endgroup

\end{document}

%% file: aastex_journal_defs.tex
\newcommand\aap{A\&A}%
\newcommand\jcap{J. Cosmology Astropart. Phys.\ }%
\newcommand\nat{Nature}%

%% file: 1IntroSHORT.tex
\section{Introduction}

The \fermi Large Area Telescope (LAT) \cite{Atwood:2009ez} provides a unique potential to measure gamma-ray intensities with an almost uniform full-sky coverage in the energy range from 20 MeV to greater than 300 GeV. Due to its good angular resolution, more than 1800 gamma-ray point sources have been reported in the second source catalog (2FGL) \cite{2012ApJS..199...31N} and more than 500 sources with a hard spectrum above 10 GeV have also been identified in the high-energy 1FHL catalog \cite{TheFermi-LAT:2013xza}. Most of these are of extragalactic origin. In addition, the excellent discrimination between charged particles and gamma rays allows LAT to directly measure diffuse gamma-ray emissions too. Note that this emission is notoriously hard to measure with Cherenkov telescopes from the ground at higher energies (above 100 GeV)  due to isotropic cosmic-ray (CR) backgrounds (for a recent effort see \cite{::2014sla}). As a result, the \fermi LAT is in a unique position to measure the diffuse emission from the Milky Way with good angular resolution and to establish an isotropic emission, that is presumably of  extragalactic origin, at energies greater than 10 GeV ~\cite{2010PhRvL.104j1101A}. 

First detected by the SAS-2 satellite  \cite{1978ApJ...222..833F} and confirmed by EGRET \cite{1998ApJ...494..523S}, the isotropic gamma-ray background (IGRB) is what remains of the extragalactic gamma-ray background (EGB) after the contribution from the extragalactic sources detected in a given survey has been subtracted.\footnote{
	 {The EGB refers here to the high-latitude gamma-ray intensity after known diffuse Galactic contributions have been modeled and subtracted. Clearly, unmodeled Galactic components (including e.g. millisecond pulsars) would still be present in the EGB as well as in the IGRB. In particular, we will consider the Galactic DM halo with its subhalos as a possible contribution to the EGB --- in spite of the fact that this might not be considered to be an ``extragalactic'' component. Moreover, emission from detected point sources is a known contributor and therefore we consider only the IGRB for our study of potential dark matter signals.}
}
%	Since the origin of the emission from detected sources is known by definition, it is appropriate to consider only the IGRB in the search for any other components, as in this analysis to search for dark matter signals.}
The \fermi LAT collaboration has recently published a new measurement of the IGRB  \cite{EGBnew} based on 50 months of data and extending the analysis described in \cite{2010PhRvL.104j1101A} down to 100 MeV and up to 820 GeV. The aim of this paper is to use this new measurement to search for evidence of a possible contribution from Weakly Interacting Massive Particle (WIMP) annihilation. This signal depends both on cosmological aspects of the DM clustering and the WIMP properties, and therefore potentially encodes a wealth of information.  With the new measurements presented in ref.~\cite{EGBnew}, it is possible to test DM models over a wide mass range, thereby testing candidates up to several tens of TeV (for \fermi LAT works that used other gamma-ray measurements for indirect DM searches see \eg,~\cite{Ackermann:2012rg,2013arXiv1310.0828T,2015arXiv150302641F}).

It is common to consider any DM annihilation signal viewed from the Sun's position as having three contributions with distinct morphological characteristics and spectra:  Galactic smooth DM distribution, Galactic substructures and extragalactic {(or, equivalently, cosmological) signal}. The extragalactic DM annihilation signal is expected to be isotropic to a large degree and constitutes the main topic of our work, but we will carefully explore the relevance of the smooth Galactic and Galactic substructure components as well. This is one important addition to the methodology presented in the original \fermi LAT publication on this topic \cite{Abdo:2010dk}. In addition to that we use improved \emph{theoretical} predictions for the extragalactic DM signal, which both takes advantage of a better knowledge of the DM clustering and its cosmic evolution (section~\ref{subsec:HM}), {\it and} utilizes a complementary and novel method \cite{Serpico:2011in,Sefusatti:2014vha} to calculate the extragalactic DM annihilation rate in Fourier space (section~\ref{subsec:PS}).  Very interestingly, we find the complementary approach to agree well with the improved predictions from the traditional method (section~\ref{sec:HM&PS}). We also explore the degree of anisotropy of the DM signal that originates from Galactic substructures, and include that component consistently with the extragalactic DM emission (section~\ref{sec:GalacticDM}). Finally, we present our constraints on the total isotropic DM signal in section~\ref{sec:limits}, for which we make minimal assumptions on the isotropic signal of conventional astrophysical origin: we either assume this astrophysical emission to be negligible (and derive {\it conservative limits}) or we assume that its contribution is perfectly known and makes up the measured IGRB intensity (thereby {estimating} the {\it sensitivity reach} of the IGRB measurement to WIMP annihilation signals).
We then study the robustness of the IGRB against adding a \emph{non-isotropic} smooth Galactic DM halo signal in the IGRB derivation  procedure  of ref.~\cite{EGBnew}. Specifically, we check for consistency of the IGRB measurement with the presence of such a Galactic DM signal in section~\ref{sec:gadgetchecks}. We summarize the main results of our work in section~\ref{sec:summary}.

%% file: 2Theory.tex
\section{Theoretical predictions of the isotropic dark matter annihilation signals}\label{2Theory}

The extragalactic gamma-ray intensity $d\phi^\mathrm{DM}/dE_0$ produced in annihilations of DM particles with mass $m_{\chi}$ and self-annihilation cross section $\sigv$ over all redshifts $z$ is given by \cite{Ullio:2002pj,Taylor:2002zd,Ando:2005xg}: 
\be
\frac{d\phi^\mathrm{DM}}{dE_0}=\frac{c\,\sigv(\Omega_{\rm DM}\rho_c)^2}{8\pi\,m_{\chi}^2}\int dz \frac{e^{-\tau(E_0,z)}(1+z)^3\zeta(z)}{H(z)}
\frac{dN}{dE}\Big{|}_{E= E_0(1+z)}\,
\label{finaleq}   
\ee 
where $c$ is the speed of light, $\Omega_{\rm DM}$ is the current DM abundance measured with respect to the critical density $\rho_c$, $H(z)$ is the Hubble parameter or expansion rate, and $dN/dE$ is the spectrum of photons per DM annihilation.\footnote{We assume here that the thermally-averaged annihilation cross section is velocity independent and that DM is composed of self-conjugated particles.} The function $\tau(E,z)$ parametrizes the absorption of photons due to the extragalactic background light. The {\it flux multiplier} $\zeta(z)$, which is related to the variance of DM density in the Universe and measures the amount of DM clustering at each given redshift, is the most uncertain astrophysical quantity in this problem. It can be expressed both in real space, making use of a Halo Model (HM) approach \cite{Cooray:2002dia}, and in Fourier space by means of a Power Spectrum (PS) approach \cite{Serpico:2011in,Sefusatti:2014vha}.

In the HM framework, $\zeta(z)$ is calculated by summing the contributions to the annihilation signal from individual halos\footnote{The term `halos' refers to all types of virialized DM clumps and structures in the Universe that lead to a DM density enhancement over the background.} of mass $M$ {from all cosmic redshifts}, $\langle F (M,z)\rangle$, and for all halo masses, \ie:
\be
\zeta(z)= \frac{1}{\rho_c}\int_{\Mmin} d M  \frac{d n}{dM} M\frac{\Delta(z)}{3} \langle F (M,z)\rangle\,,\label{zetaz}
\ee
where $\Mmin$ is the minimum halo mass considered, and $\Delta(z)$ and $\frac{d n}{dM}$ are the DM halo-mass over-density and the {\em halo mass function}, respectively. The mean halo over-density $\Delta(z)$ is defined with respect to the {\it critical} density of the Universe and its value determines the virial radius of a halo, $R_{\Delta}$, at each redshift. In this paper we will adopt $\Delta(0) = 200$, following previous choices in the literature, and compute it at different redshifts as in ref.~\cite{Bryan:1997dn}. The halo mass function $\frac{d n}{dM}$ is normalized by imposing that all mass in the Universe resides {inside} halos (see \cite{Ullio:2002pj} for more details). 
$ \langle F (M,z)\rangle$ in turn depends on the DM {\em halo density profile} and the halo size. Halo density profiles are determined by N-body cosmological simulations, with the most recent results favoring cuspy Navarro-Frenk-White (NFW) \cite{Navarro:1995iw} and Einasto halos \cite{Graham:2005xx,Navarro:2008kc}, while some astrophysical observations favor cored halos, \eg, Burkert density profiles \cite{Burkert:1995yz,Salucci:2000ps}. The density profile $\kappa$ can be easily expressed in terms of a dimensionless variable $x=r/r_s$, $r_s$ being the scale radius at which the effective logarithmic slope of the profile is $-2$. In this prescription, the virial radius $R_{\Delta}$ is usually parametrized by the {\it halo concentration} $c_{\Delta}=R_{\Delta}/r_s$ and the function $F$ can be written as follows:

\be
F(M,z)\equiv c_{\Delta}^3(M,z)\frac{\int_0^{c_{\Delta}}dx\,x^2\kappa^2(x)}{\left[\int_0^{c_{\Delta}} dx\,x^2\,\kappa(x)\right]^{2}}\,.\label{FMz}
\ee
More realistically $F$ is an average over the probability distribution of the relevant parameters (most notably $c_{\Delta}$). Note that the above expression depends on the third power of the concentration parameter. From simulations it has been determined that the halo mass function and halo concentration decrease with halo mass and, consequently, the flux multiplier $\zeta(z)$ given by eq.~(\ref{zetaz}) is dominated by small mass halos  {(which could be either field halos or subhalos\footnote{Subhalos are halos within the radius of another halo. A halo that does not reside inside any other halo will be referred to as a field halo or, simply, halo.}; see section~\ref{subsec:HM} for further discussion}). Furthermore, DM halos contain populations of subhalos, possibly characterized by different mean values of the relevant parameters (\eg, different concentrations than those of field halos). The signals from subhalos are typically included by adding an extra term in eq.~(\ref{zetaz}) to account for halo substructure, see \cite{Ullio:2002pj}.

As noted in  \cite{Serpico:2011in} the flux multiplier can also be expressed directly in terms of the {\it non-linear} matter power spectrum $P_{NL}$ (the Fourier transform of the two-point correlation function of the matter density field):%}):
\be
\zeta(z)\equiv \langle \delta^2 (z)\rangle= \int ^{\kmax} \frac{d\,k}{k}\frac{k^3 P_{NL}(k,z)}{2\pi^2}\equiv\int ^{\kmax} \frac{d\,k}{k}\Delta_{NL}(k,z) \label{zetazNL},
\ee 
where $\Delta_{NL}(k,z) \equiv k^3P_{NL}(k,z)/(2\pi^2)$ is the dimensionless non-linear power spectrum and $\kmax(z)$ is the scale of the smallest structures that significantly contribute to the cosmological annihilation signal. Loosely speaking, one could define a relation $M~=4/3 \pi \rho_h\; (\pi/k)^3$ with $\rho_h$ the characteristic  density of the DM halo. Therefore $\kmax$ is the PS counterpart to the minimal halo mass $\Mmin$ in eq.~(\ref{zetaz}) in the HM prescription.

The extrapolation to masses or $k$ scales beyond the resolution of N-body simulations is the largest source of uncertainty in predictions of the extragalactic signal of DM annihilation, since the smallest scales expected for the WIMP models are far from being probed either by astrophysical observations or simulations.\footnote{Notable exceptions on the simulation side are the works \cite{2005Natur.433..389D,2013JCAP...04..009A}, which simulated a few individual $\sim10^{-6} \Msun$ halos and the recent work  \cite{2014ApJ...788...27I}, where for the first time dozens to thousands of halos have been resolved with superb mass resolution in the range $2\times10^{-6} - 4\times10^{-4} \Msun$.  {Note, though, that these simulations were performed at high redshifts, so some extrapolation is needed to describe structures at the present time.}} Thus, different methods of extrapolating to the smallest masses can lead to completely different results for the relevant quantities. Typical expectations for the minimum halo masses in WIMP models are in the range $\Mmin\in [10^{-9}, 10^{-4}]\,\Msun$ (see~\cite{Green:2005fa,Bringmann:2009vf,Cornell:2013rza} and refs. therein), while we only have observational evidence of structures down to $10^7\,\Msun$~\cite{Geha:2008zr} implying that extrapolations are required to span $\gsi 10$ orders of magnitude in halo mass (or $\gsi 3$ orders of magnitude in $k$). 

Both ways of expressing $\zeta$, eq.~(\ref{zetaz}) and (\ref{zetazNL}), have advantages and disadvantages. While eq.~(\ref{zetaz}) is given in real space and thus deals with `intuitive' quantities, it depends to a large extent on several poorly constrained parameters, most notably the concentration and halo mass function. This is particularly true for the smallest halos, which, as we have noted, are expected to dominate in the evaluation of $\zeta$. The same is applicable to the subhalo population, whose internal properties and abundance are even less well understood. In contrast, eq.~(\ref{zetazNL}) depends only on one quantity directly determined from N-body simulations,\footnote{The quantity is the non-linear matter power spectrum, which is determined using only a matter density map, without invoking the concept of halos and without relying on standard halo finders.} which can be extrapolated based on simple scale-invariant arguments, but lacks the intuitive interpretation of breaking structures into individual halos and subhalos, relevant when comparing the expected signals from Milky Way substructures with the total cosmological one. 

In this work, we use both of these approaches in parallel: the HM is used to define our \emph{benchmark model} following simple but well motivated arguments for the choice of the relevant ingredients, and the PS framework is used to calculate the associated \emph{uncertainty} due to extrapolation to small (unresolved) scales. 

%% file: 21HaloModel.tex
\subsection{Halo-model setup}\label{subsec:HM}

The DM annihilation signal from a halo depends on the third power of the concentration (see eq.~\ref{FMz}), and the results will be extremely sensitive to the way the concentration-mass relation is extrapolated to low mass. A common practice in the past has been to use a single power law of $M$ for $c(M)$ all the way down to the minimum halo mass (see \eg \cite{Zavala:2009zr,2011PhRvD..84l3509P,2012MNRAS.419.1721G}). In most cases, these power-law extrapolations assign very high concentrations to the smallest halos, resulting in very high DM annihilation rates. In addition, these results are extremely sensitive to the power-law index used. However, these power-law extrapolations are unphysical and not expected in the $\Lambda$CDM cosmology \cite{2014MNRAS.442.2271S}. Indeed, since natal CDM concentrations are set by the halo formation epoch and the smaller structures collapse at nearly the same time in the early Universe, low-mass halos are expected to possess similar natal concentrations, and therefore are expected to exhibit similar concentrations at the present epoch as well. In other words, the expectation that $c(M)$ at the small mass end flattens is deeply rooted in the  $\Lambda$CDM framework. This kind of behavior is correctly predicted by the $c(M)$ model of ref.~\cite{2012MNRAS.423.3018P}, which explicitly relates halo concentrations to the root mean square ($r.m.s.$) of the matter density fluctuations, $\sigma(M)$. In addition, both the $\Lambda$CDM expectations and the model predictions at the low-mass end are supported by the (few) results that come from N-body simulations that were specifically designed to shed light on this extreme small-halo-mass regime \cite{2005Natur.433..389D,2013JCAP...04..009A,2014ApJ...788...27I}. This was also pointed out in ref. \cite{2014MNRAS.442.2271S} where, making use of all available N-body simulation data, the authors examine the $c(M)$ relation at redshift zero for all halo masses (\ie from Earth-mass microhalos up to galaxy clusters). We refer the reader to these works for further details and discussion on the $c(M)$ behavior at the low mass end.  {Note, though, that while the flattening of $c(M)$ at lower masses is well motivated by $\Lambda$CDM and supported by simulations, the evolution of low-mass halos and subhalos from their formation times to the present and, in particular, their survival probability, is far from being completely verified in simulations or in analytical work (see, e.g., Refs.~\cite{Zhao:2005mb,Berezinsky:2005py,Goerdt:2006hp}). Thus, even if the actual DM particle properties were known, unknowns in the subhalo survival probabilities and the impact of baryonic-feedback processes on the dark sector would still contribute to significant uncertainties on the theoretical expected DM signal. } 

\smallskip
In what follows we define our benchmark halo model.
\begin{itemize}
\item{Cosmological parameters:}
 we assume those recently derived from \planck data \cite{2013arXiv1303.5076P}, \ie, $\Omega_M=0.315$, $h=0.673$, $\sigma_8=0.834$.\footnote{The $\sigma_8$ is the $r.m.s.$ amplitude of linear matter fluctuations in 8 h$^{-1}$ Mpc spheres at $z$ = 0 with the Hubble constant $h$ defined via $H(z=0) = h \times 100$~km/s/Mpc.} 
\item{Minimum halo mass:}
 here we make a choice that has become standard in the DM community, \ie $\Mmin=10^{-6}$  h$^{-1}$ $\Msun$, though we caution the reader that in supersymmetric models this value is simply expected to be in the $[10^{-9}, 10^{-4}]\,\Msun$ range \cite{Green:2005fa,2006PhRvL..97c1301P,Bringmann:2009vf,Cornell:2013rza}. {We explore how predictions of the extragalactic DM signal are affected by the adopted $\Mmin$ value later in Section~\ref{sec:HM&PS}.}
\item{Halo concentration:}
 we adopt the model of ref.~\cite{2012MNRAS.423.3018P} discussed above, which implicitly assumes a critical over-density, $\Delta$ in eq.~(\ref{zetaz}), equal to 200; thus it gives a relation between $c_{200}$ and $M_{200}$.
\item{Halo mass function:}
 we use the state-of-the-art halo mass function as proposed by Tinker {\it et al.}, \cite{2008ApJ...688..709T}, but with the parameters deduced by \cite{2012MNRAS.423.3018P} at redshift zero for the \planck cosmology.\footnote{We still respect the parameters' z-dependence found in ref.~\cite{2008ApJ...688..709T} though.}  Inspired by the prior work by Sheth \& Tormen \cite{2002MNRAS.329...61S}, the Tinker {\it et al.} function gives a better agreement with N-body simulations especially at the high-mass end.
\item{DM density profile:}
 we use the familiar NFW profile \cite{Navarro:1995iw}. The predicted values of the flux multiplier $\zeta(z)$ are fairly independent of this choice (for the standard assumptions of the profiles considered in this work). It was shown for example, that the quantity $ \langle F (M,z)\rangle$ in eq.~(\ref{FMz}) changes at the $10\%$ level when the profile is changed from the NFW profile to the cored Burkert profile (see also figure~3 in ref.~\cite{Ullio:2002pj}).
\item{Contribution of the subhalo population:}
 while both the halo mass function and halo concentrations are reasonably well studied, the properties of the subhalos are more uncertain. In order to estimate the DM annihilation rate produced in subhalos and its contribution to the total extragalactic signal, here we resort to the results of the study recently presented in ref.~\cite{2014MNRAS.442.2271S}. Two parameters that control the amount of substructure and therefore its contribution to the annihilation flux are the minimum subhalo mass and the slope of the subhalo mass function. Following our choice for main DM halos, we adopt a value of $10^{-6}$~h$^{-1}\Msun$ for the first of these two ingredients. As for the slope of the subhalo mass function, $\frac{d n_{sub}}{dM} \propto (m_{sub}/M_{host})^{-\alpha}$, we take $\alpha = 2$, although we will also examine the impact of changing this exponent to $1.9$ in the next section.
Following results found in high-resolution N-body cosmological simulations of Milky-Way-sized halos above their mass resolution limits \cite{Diemand:2006ik,Springel:2008cc}, the normalization of this subhalo mass function is such that the mass contained in subhalos down to $10^{-6}$~h$^{-1}\Msun$ is $\sim 45$\% of the total parent halo mass.\footnote{Note, however, that the substructure mass fraction which is {\it resolved} in the simulations is only of about 10\%. The 45\% quoted in the text refers to the total (resolved plus unresolved) fraction which is obtained by extrapolating the subhalo mass function down to $10^{-6}$  h$^{-1}\Msun$.} 
The mentioned values correspond to the reference substructure boost model in ref.~\cite{2014MNRAS.442.2271S}. We note that this model implicitly assumes that both subhalos and field halos share the exact same internal properties, which is probably not the case. Nevertheless, as discussed in ref.~\cite{2014MNRAS.442.2271S}, this choice represents a conservative case in terms of expected  gamma-ray intensity from annihilations in subhalos.  
\end{itemize}

We show the  extragalactic intensity predicted by the benchmark HM described above in figure~\ref{fig:PS_HM_comp}. Since the uncertainty of this signal is not easy to quantify within the HM approach given its dependence on the many variables involved (see, \eg,~\cite{Zavala:2009zr,2013MNRAS.429.1529F,Mack:2013bja,Ng:2013xha}), we use the PS approach to define our uncertainty band, as will be detailed in the next subsection.

%% file: 22PowerSpectrum.tex
\subsection{Power-spectrum setup}\label{subsec:PS}

In this section we estimate the uncertainty affecting the evaluation of the flux multiplier, $\zeta$. We focus on the PS approach, \ie eq.~(\ref{zetazNL}), which relies on the work done in ref.~\cite{Sefusatti:2014vha} using the data from the Millenium-II N-body cosmological simulation (MS-II) \cite{BoylanKolchin:2009an}. MS-II has the highest mass resolution among available large-scale structure simulations, but its cosmological parameters are from early \wmap 1-year \cite{Spergel:2003cb} and 2dF Galaxy Redshift Survey \cite{Colless:2000et} data. 
In ref.~\cite{Sefusatti:2014vha}, it was shown that the largest uncertainties in the calculation of $\zeta$ come from modeling the behavior of the power spectrum beyond the MS-II resolution and the exact value of the cut-off scale. Lacking a better theory, in \cite{Serpico:2011in,Sefusatti:2014vha} the extrapolation of the data was inspired by the behavior of $\Delta_{NL}(k)$ found in the simulation itself.  
In particular, {\it at redshift zero}, it was assumed that the true $\Delta_{NL}(k)$ value is bracketed by two alternative extrapolations for scales smaller than $k > k_{1\%}$:
\begin{eqnarray}
\Delta^\mathrm{min}(k)&=&\Delta_{NL}(k_{1\%}) \label{extrap0}  \label{eq:PSmin} \\ 
\Delta^\mathrm{max}(k) &=& \Delta_{NL}(k_{1\%})\left(\frac{k}{k_{1\%}}\right)^{n_{\rm eff}} \label{extrap1} \,
\end{eqnarray}
where $n_{\rm eff}={d\ln\Delta(k_{1\%})}/{d\ln k}$ and $k_{1\%}$ is the scale at which the shot noise contribution to the power spectrum $\Delta_{NL}$ is $1\%$ and sets the resolution threshold. In other words, $\Delta^\mathrm{max}(k)$ was found by imposing that the spectral index $n_{\rm eff}$ found at the resolution threshold stays constant at larger $k$ scales. This is conservative, as a flattening of the power spectrum is predicted by theoretical arguments (see, \eg,~\cite{Sefusatti:2014vha} for further details) while the extrapolation is from the trend observed in simulations at the smallest resolved mass scales.  {The clustering properties and the central parts of the DM density profiles at the smallest scales are not directly measured quantities and extrapolations are thus made under the assumptions mentioned above.}

At {\it higher redshifts}, where MS-II resolves only the largest co-moving scales, an additional constraint was imposed on $\Delta^\mathrm{max}(k)$: the ratio of the nonlinear to linear power spectrum is not allowed to decrease with time at any fixed scale $k$ (\ie, the Universe can only become clumpier in time, at every co-moving scale). Similarly, in order to also constrain the {\it minimal} $\zeta$ value more precisely, $\Delta^\mathrm{min}(k)$ is required to have an effective spectral index bounded from below by the spectral index predicted in linear theory.  For details on the physical motivations for these extrapolations see \cite{Sefusatti:2014vha}.

The cosmological parameters used for the MS-II simulation differ significantly from the values recently measured by \planck, which we adopted in the HM approach. The most critical factor in this regard is probably the value of $\sigma_8$, since this parameter has a large influence over the growth of fluctuations in the early Universe and thus on the subsequent evolution of structures. The results of the HM approach are derived assuming the most-up-to-date value $\sigma _8=0.835$ recently given by \planck data, but MS-II adopts $\sigma _8=0.9$.
Thus, in order to make a fair comparison between the two methods, we follow \cite{Sefusatti:2014vha} to apply the \planck cosmology to  the results derived from the MS-II. 
We comment on the particular choice of the cut-off scale, $\kmax$, in the next subsection.

%% file: 23HM_PS_Comparison.tex
\subsection{Comparison of the two approaches and their cosmological dark matter signal predictions} \label{sec:HM&PS}

\begin{figure}[t]
\centering
\includegraphics[width=0.75\textwidth]{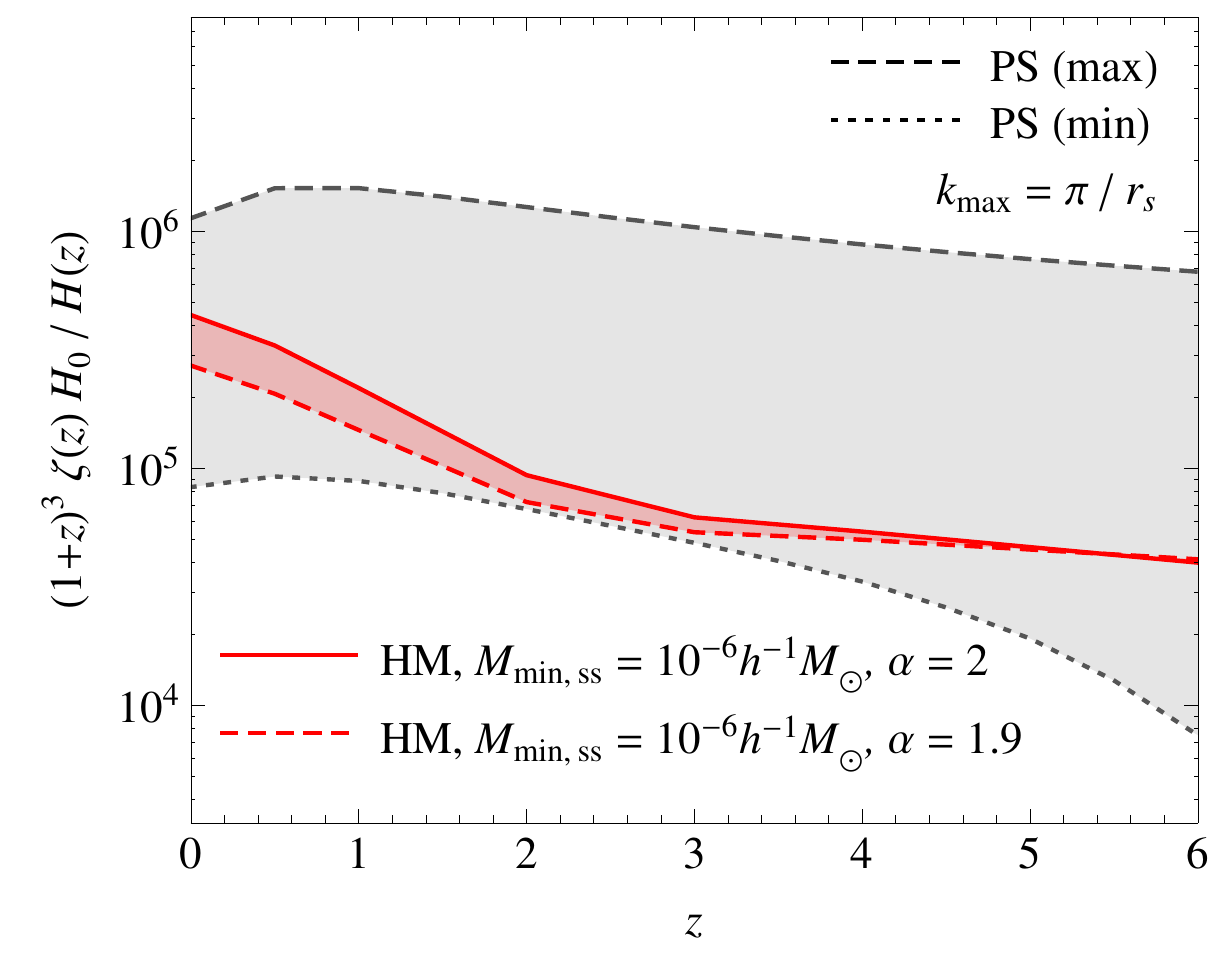}
\caption{Normalized $\zeta$ as a function of redshift. A value of $\Mmin=10^{-6}$ h$^{-1}\Msun$ was used in both the PS (gray) and HM predictions (red). The benchmark HM model detailed in section~\ref{subsec:HM} is shown by the red solid line. The red dashed line corresponds to the case in which the slope of the subhalo mass function varies from the fiducial $\alpha=2$ to $1.9$ (\ie, less substructure). The dotted line, labeled `PS (min)', shows the minimum approximation from Equation \ref{eq:PSmin} in the PS approach, while the dashed line, `PS (max)', shows the maximum approximation given by Equation \ref{extrap1}.}%\masc{i) maybe include the HM with no substructures? (see text too); 
\label{fig:PS_HM_comp}
\end{figure}

In figure~\ref{fig:PS_HM_comp} we compare the values of $\zeta$ obtained by means of the HM and PS approaches described above. In order to make a proper comparison and extract meaningful conclusions, we call attention to a few caveats.
%
%%%%%%%%%%%%%%%%%%%%%%%%%%%%%%%%%%%%%%%%%%%%%%%%%%%%%%%%%%%%%%%%%%%%%%%%
%%%%%%%%%%%%%%%%%%%%%%%%%%%%%%%%%%%%%%%%%%%%%%%%%%%%%%%%%%%%%%%%%%%%%%%%

A conceptual issue in comparing the results of the two approaches is the determination of their cut-off scales. Such a cut-off exists, for example, due to the free streaming length of DM particles after kinetic decoupling \cite{Green:2005fa,Bringmann:2009vf}, which gives the highest frequency relevant Fourier mode $\kmax$.\footnote{In the linear regime, the quantity $\kmax$ is a redshift independent quantity.} In the PS approach, $\kmax$ sets a sharp upper cut-off on the matter power spectrum at the scale corresponding to the presumed free streaming length in the linear regime, below which structures do not contribute to the DM annihilation signal. In the HM approach, the cut-off is instead imposed as a minimum halo mass,\footnote{Chosen to correspond to the linear free streaming scale.} below which no halos are formed and therefore no annihilation signal is expected. Within each DM halo, however, the annihilation signal is calculated by extrapolating the adopted DM density profile of the halo down to $r \rightarrow 0$. Strictly speaking the HM thus includes Fourier modes all the way to infinity, even if the largest wave numbers would not contribute much. Note that the exact shape of the DM density profile at those small scales as well as the mass of the first virialized objects to form is still scarcely probed in simulations. 
As described in section~\ref{subsec:HM} and \ref{subsec:PS}, we chose $\Mmin=10^{-6}$ h$^{-1}\Msun$ as the cut-off scale in our benchmark HM model and we adopt $\kmax=\pi/r_s$ as the default choice for the corresponding cut-off in the case of the PS approach (where $r_s$ is derived assuming the $\Delta=200$ value used in the HM approach, {\it i.e.}\ $r_s = R_{\Delta}(z) c_{\Delta}(z)=R_{200}(z) c_{200}(z)$). We recall that other motivated choices for $\kmax$ are possible, and refer the reader to ref.~\cite{Sefusatti:2014vha} for further details.

\begin{figure}[t]
\centering
\includegraphics[width=0.7\textwidth]{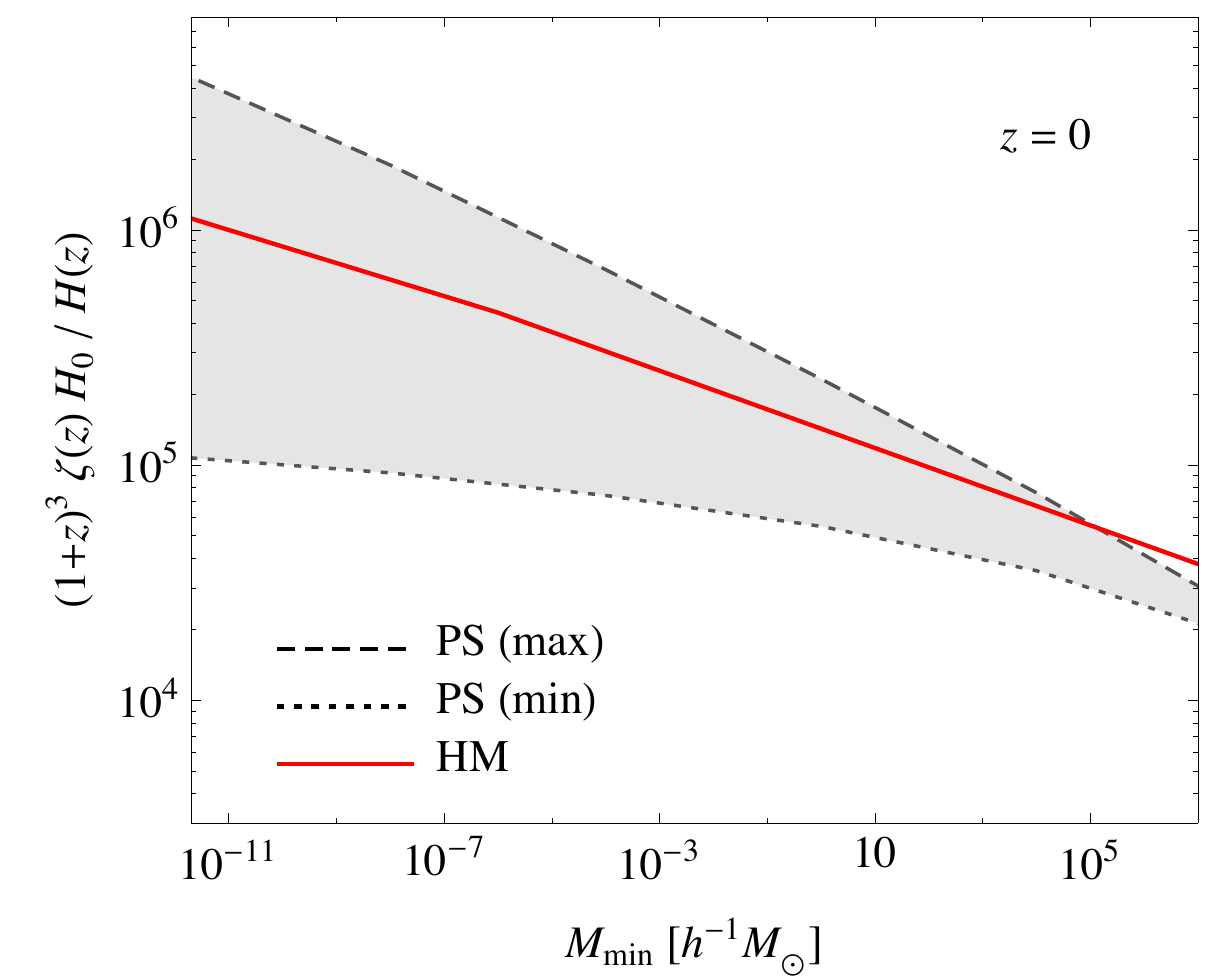}
\caption{Variation of the value of $\zeta$ at $z=0$ as a function of the minimum halo mass considered, $\Mmin$. Dotted and dashed lines represent minimum and maximum approximations in the PS approach as described by Eqs. (\ref{extrap0}) and (\ref{extrap1}), respectively. {In this work, we adopt $\Mmin=10^{-6}$  h$^{-1}$ $\Msun$ as our fiducial value.} See text for further details. %\masc{$\Mmin$ in the x-axis instead of M would be better.}
}
\label{fig:PS_zeta2_mcut}
\end{figure}

Another caveat when comparing HM to PS in figure~\ref{fig:PS_HM_comp} is the signal contribution from substructures that are present in extragalactic DM halos. As discussed in the previous section, the PS-derived $\zeta$ values implicitly include such signal by covering contributions down to length scales $\sim\pi/\kmax$. In the HM approach, by contrast, the substructures' contribution is calculated by introducing additional parameters. We show in figure~\ref{fig:PS_HM_comp} the HM prediction for two different scenarios: the ones corresponding to the minimum and maximum allowed values of the (substructure-induced) boost factor to the annihilation signal from field halos as predicted in ref.~\cite{2014MNRAS.442.2271S} for a fixed value of $M_{\rm min,ss}=10^{-6}$ h$^{-1}\Msun$. In this case, the differences in boost factors are due to different assumptions for the slope of the subhalo mass function, $\alpha$ (larger $\alpha$ values lead to more substructure and thus to larger boosts). As a consequence of the aforementioned limitations, we expect some uncertainty when making a quantitative comparison between the HM and PS approaches. Nevertheless, the agreement is quite good as can be seen in figure~\ref{fig:PS_HM_comp}, our benchmark HM prediction being within the minimum and maximum PS values at all redshifts. 
\medskip
We have so far explored the expected WIMP signal for a given assumed cut-off scale $\Mmin$ (or, equivalently, $\kmax(z)$ defined by $\pi/r_s$). However, this effective cut-off scale can vary significantly between various DM candidates, depending for example on their free-streaming lengths, as discussed in, \eg, \cite{Bringmann:2009vf}. In figure~\ref{fig:PS_zeta2_mcut} we explore this dependence of $\zeta$ on the cut-off scale $\Mmin$ and $\kmax(z)$. 
%
%%%%%%%%%%%%%%%%%%%%%%%%%%%%%%%%%%%%%%%%%%%%%%%%%%%%%%%%%%%%%%%%%%%%%%%%
%%%%%%%%%%%%%%%%%%%%%%%%%%%%%%%%%%%%%%%%%%%%%%%%%%%%%%%%%%%%%%%%%%%%%%%%
%

\begin{figure}[t]
\centering
\includegraphics[width=0.7\textwidth]{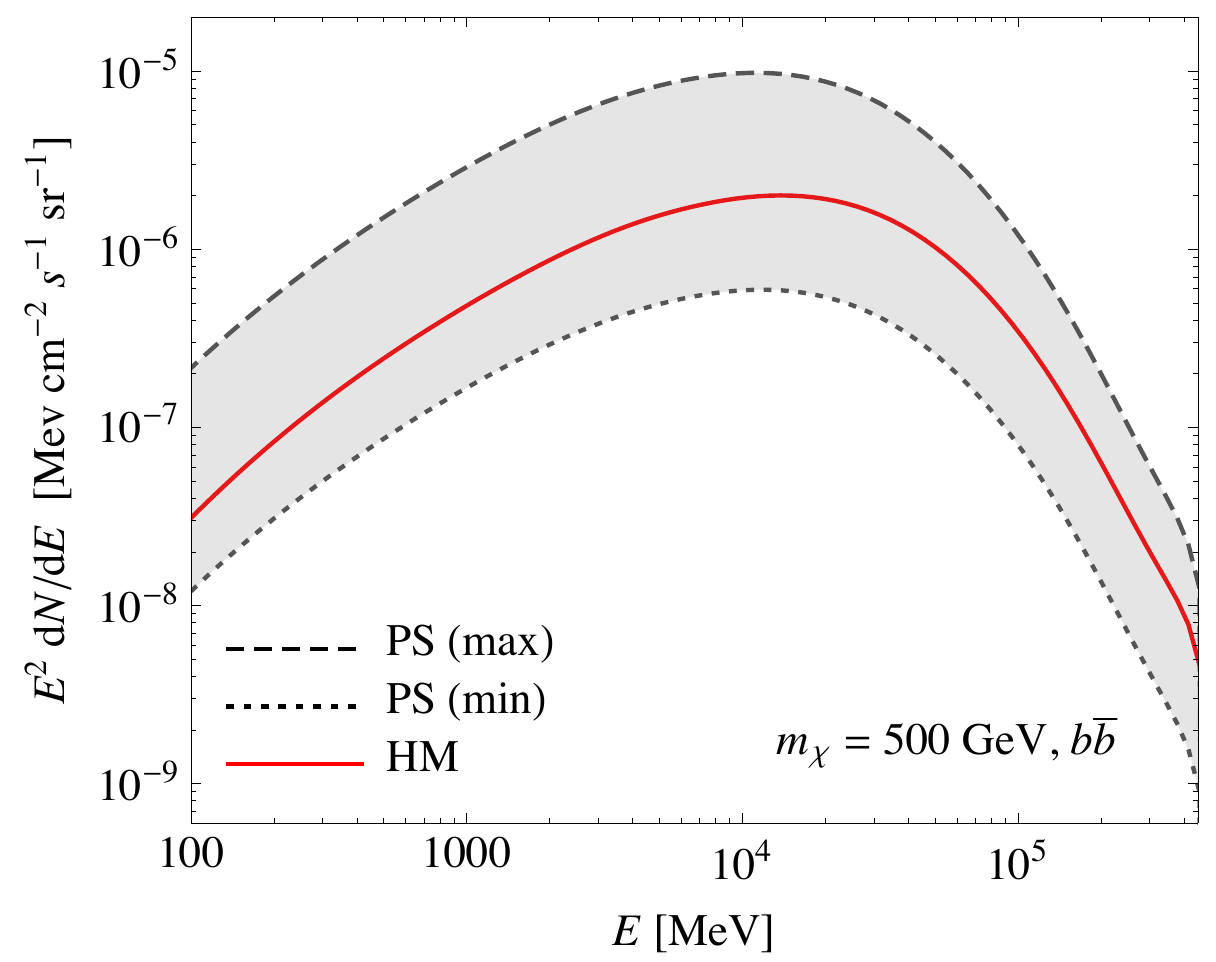}
\caption{Comparison of the predicted cosmological DM-annihilation-induced gamma-ray intensities as given by both the PS and HM approaches. The $\zeta$ values implicitly used are those given in figure~\ref{fig:PS_HM_comp}. This particular example is for a 500 GeV DM particle annihilating to $b{\bar b}$ channel with a cross section $\langle \sigma v\rangle = 3\times 10^{-26}$cm$^3$/s. The signal range shown in gray (which is computed within the PS approach) translates directly into uncertainties in the DM limits of section~\ref{sec:limits}.}
\label{fig:PS_HM_comp_Flux_band}
\end{figure}

\medskip
Finally, we now turn to the  calculation of the extragalactic gamma-ray spectrum, by integrating $\zeta (z)$ over all redshifts and folding it with the induced spectrum from WIMP annihilations,  see eq.~(\ref{finaleq}). To model the attenuation of gamma rays traveling through cosmological distances, we adopt the Dom\'inguez {\it et al.} model \cite{2011MNRAS.410.2556D} for the extragalactic background light (EBL), which represents the state-of-the-art and is fully consistent with the first direct detections of the EBL attenuation by the %\fermi 
LAT \cite{2012Sci...338.1190A} and H.E.S.S. collaborations \cite{2013A&A...550A...4H}, and with the recently measured cosmic gamma-ray horizon \cite{2013ApJ...770...77D}. In figure~\ref{fig:PS_HM_comp_Flux_band} we show a typical example for the gamma-ray flux resulting from 500 GeV DM particles annihilating through the $b{\bar b}$ channel. 
%
%%%%%%%%%%%%%%%%%%%%%%%%%%%%%%%%%%%%%%%%%%%%%%%%%%%%%%%%%%%%%%%%%%%%%%%%
%%%%%%%%%%%%%%%%%%%%%%%%%%%%%%%%%%%%%%%%%%%%%%%%%%%%%%%%%%%%%%%%%%%%%%%%
%
As we will discuss in detail in section~\ref{sec:limits}, the gray band in this figure translates directly into an uncertainty in the DM limits.

%% file: 24GalacticSS.tex
\subsection{Galactic dark matter signal contributions}    \label{sec:GalacticDM}

Within a DM halo, the DM distribution has two distinct components: a centrally-concentrated smooth distribution and a population of subhalos. While the smooth component has been studied extensively over a large span of halo masses, it is only more recently that  several high-resolution simulations of Milky-Way-size halos addressed the subhalo component in more detail. These simulations quantified the radial distribution of subhalos inside the host halo, their abundance and their overall luminosity \cite{Diemand:2006ik,Springel:2008by,2009MNRAS.398L..21S}. The smooth and subhalo components are fundamentally different in terms of the level of anisotropy and, in the following, we quantify their relevance for this work.

The smooth DM density profile of the main halo, $\rho(r)$, is found in pure DM N-body simulations to be NFW \cite{Navarro:1995iw} or Einasto \cite{1968PTarO..36..414E,2004MNRAS.349.1039N}, while more computationally-expensive simulations which include baryonic effects have not yet converged on the profile shape in the inner regions of the Galaxy, finding both more cored \cite{2012ApJ...744L...9M,2013ApJ...765...10K} and more cuspy profiles \cite{Gondolo:1999ef,Gustafsson:2006gr}. Indeed, the solution to this issue might be substantially more complicated: the ability for galaxies to retain their DM cusps may depend on the ratio of their stellar and halo masses \cite{2013arXiv1306.0898D}. Observational tracers of the gravitational potential also cannot be used to determine the DM profile within $\lsi$ 2 kpc from the Galactic center as it is gravitationally dominated by baryons \cite{Iocco:2011jz,Nesti:2013uwa}. However, the considerable uncertainties in the profile shape of the inner Galaxy are not critical for this work. We will deal with Galactic latitudes $\gsi$ 20$^\circ$, $\sim 3$ kpc from the Galactic Center, a region in which simulations and astrophysical evidence converge on the $\rho \sim r^{-2}$ profile behavior.

Once the radial profile of the DM density is fixed, the remaining uncertainty lies in its overall normalization, \ie the value of $\rho_0$ at the solar radius, {which can take values in the range $\rho_0=0.2$--$0.8$ ${\rm GeV\,cm^{-3}}$ \cite{Cirelli:2010xx,Nesti:2013uwa}}.\footnote{In principle the values of $r_s$ and $\rho_0$ are not independently constrained, and they should be correlated consistently. However, for our purposes the asserted ranges represent a fair description of the uncertainties expected for the DM signal.} The gamma-ray signal is proportional to the square of $\rho$ and therefore its uncertainty becomes greater than an order of magnitude in the worst case. We will {\it not} consider any (portion) of the signal from the smooth Galactic DM distribution to contribute to the isotropic emission (as has been done in some previous works, \eg\ \cite{2010JCAP...11..041A}). Instead, we find that the whole-sky DM template can be partially degenerate with at least one of the astrophysical components present in the Galactic foreground emission. We will treat this signal from the smooth Galactic DM halo as an additional component of the foreground Galactic diffuse emission instead. We further discuss this choice and its impact on the derivation of the IGRB spectra in section~\ref{sec:gadgetchecks}.

As far as the diffuse gamma-ray intensity from DM annihilations in the Galactic subhalo population is concerned, some earlier works \cite{Pieri:2009je,Abdo:2010dk} found that it could appear isotropic in our region of interest, and we explore this issue here in more detail. The exact {distribution} in Galactocentric distance  of subhalos is currently not well determined: in the Via Lactea~II simulation \cite{Diemand:2006ik} the subhalos follow the so-called anti-biased relation with respect to the smooth DM density profile, \ie $\rho_{\rm sub} (r) \propto r \times \rho_{\rm NFW}/(r + r_a)$ with $r_a \simeq 85.5$ kpc, while in the Aquarius simulation \cite{Springel:2008cc,Springel:2008by} they can be described by $\rho_{\rm sub} \propto \rho_{\rm Einasto} \propto \exp \left\{-\frac{2}{\alpha_E}\left[\left(\frac{r}{r_s}\right)^{\alpha_E} - 1\right] \right\}$, with a particularly large scale radius $r_{s} = 199$~kpc and $\alpha_E = 0.678$ \cite{Pieri:2009je}.  

\begin{figure}[t]
\hspace{-0.2cm}
\includegraphics[width=0.55\textwidth]{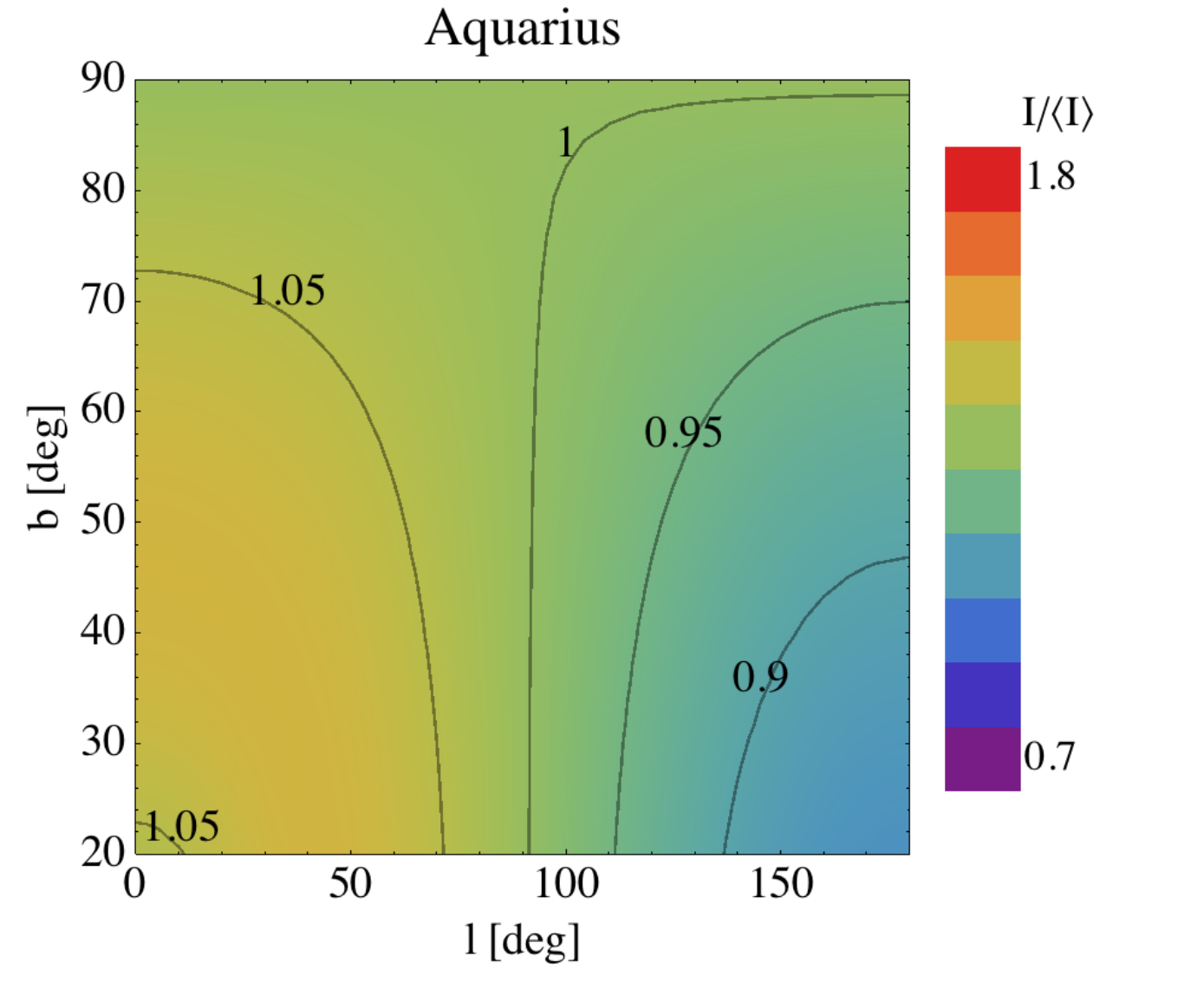}
\hspace{-0.8cm}
\includegraphics[width=0.55\textwidth]{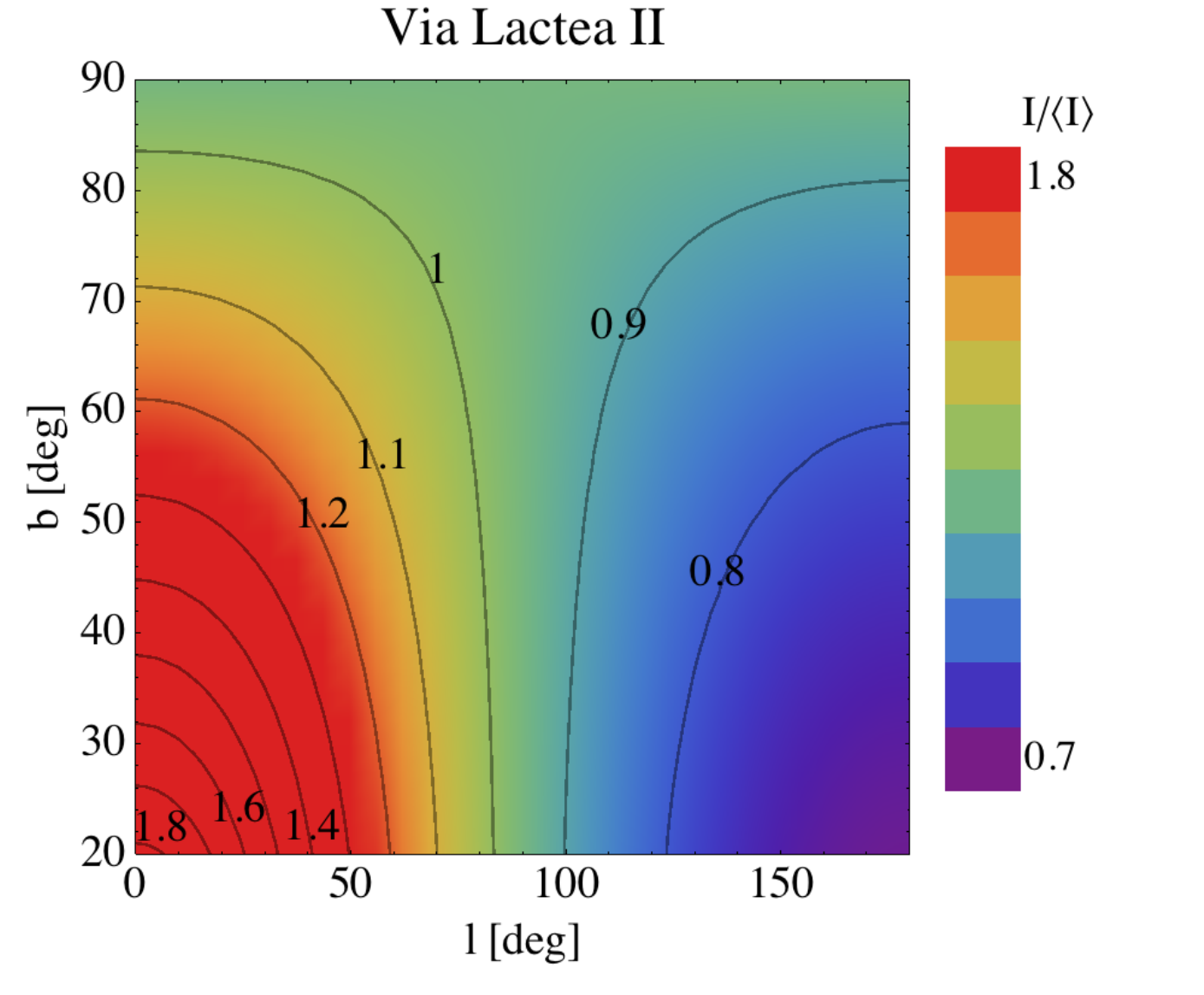}
\caption{Anisotropy in the gamma-ray annihilation signal from the subhalo distribution found in Aquarius  \cite{Springel:2008cc} (left) and Via Lactea II \cite{Diemand:2006ik} (right) simulations, with the former following the prescription in ref.~\cite{Pieri:2009je}. The plots show the intensities of the substructures relative to their average intensity in the $|b|>20^\circ$ region.
} \label{fig:SS_aniso}
\end{figure}

In figure~\ref{fig:SS_aniso} (left) we show that the substructures give on average $\lsi 10$ \% anisotropy in the relative intensity $I/\langle I\rangle$ for a DM-annihilation-induced signal when the DM substructure distribution is described as in the original {Aquarius} simulation paper \cite{Springel:2008cc}. However, for the same Aquarius simulation, authors in ref.~\cite{Pieri:2009je} find a signal from the substructures that is less isotropic.

Yet, the variations with respect to the average intensity are significantly less than a factor of 2 for latitudes $|b|> 20^\circ$. From the {Via Lactea II} simulation, the results in ref.~\cite{Diemand:2006ik} give an anisotropy of $I/\langle I\rangle$ that is also less than a factor of 2, see figure~\ref{fig:SS_aniso} (right). 
These numbers can be compared to the Galactic smooth DM annihilation signal intensity that varies by more than a factor 30 for latitudes $|b|> 20^\circ$ for the NFW profile that we used.
%
%%%%%%%%%%%%%%%%%%%%%%%%%%%%%%%%%%%%%%%%%%%%%%%%%%%%%%%%%%%%%%%%%%%%%%%%
%%%%%%%%%%%%%%%%%%%%%%%%%%%%%%%%%%%%%%%%%%%%%%%%%%%%%%%%%%%%%%%%%%%%%%%%
%

We compare these findings with the sky residuals found when deriving the IGRB spectrum in ref.~\cite{EGBnew}, which are at the $\lsi 20\%$ level. We conclude that, at least in the case of the Aquarius simulation, the Galactic substructure would lead to a sufficiently isotropic signal and thus we add it to the extragalactic signal when setting the DM limits.

However, while the spatial distribution of the DM signal is taken directly from the aforementioned simulations, the total signal strength is assigned self-consistently with the total substructure signal as of a typical Milky-Way-sized DM halo (as used in the calculation of the extragalactic signal within the HM approach). On the other hand, while the extragalactic DM signal comes from a large ensemble of halos, the properties of the Galactic substructure population are determined by the particular formation history of the Milky Way galaxy. In that sense, its properties do not have to correspond to the mean values found throughout the Universe, and in fact the Milky Way is found to be atypical in several respects \cite{BoylanKolchin:2009an}.  For that reason, we consider two DM substructure prescriptions that introduce boosts of the total annihilation rate in our Galactic DM halo by factors of 3 and 15. This range follows from the prescription of \cite{2014MNRAS.442.2271S} using a fixed minimum subhalo mass of $M_{\rm min,ss}=10^{-6}$ h$^{-1}\Msun$ but varying the slope of the subhalo mass function ($\alpha=1.9$ and $2$, respectively). 
In the next section, we will show limits on DM annihilation cross sections from assuming these two bracketing values on the substructure boost.
Changing $M_{\rm min,ss}$ to, \eg, $10^{-12}$ h$^{-1}\Msun$ would not affect the lower boost factor, but would increase the upper boost factor bound from 15 to about 40 (see ref.~\cite{2014MNRAS.442.2271S}).

Some of the largest or closest Galactic DM substructures could eventually be resolved as discrete gamma-ray sources. The contribution from few individual subhalos to the total isotropic WIMP signal is not substantial, but nonetheless current constraints on DM signals from dwarf spheroidal galaxies \cite{2013arXiv1310.0828T,2015arXiv150302641F}, as well as the non-detection of DM signals from unassociated gamma-ray sources, \eg, \cite{2012ApJ...747..121A, 2012JCAP...11..050Z, 2014PhRvD..89a6014B}, significantly limit the total annihilation signal from the DM subhalos in the Milky Way. Our approach is to include the total expected DM signals from all subhalos of all masses in our evaluation of the DM signal contribution to the IGRB, but when DM limits from, \eg, dwarf spheroidal galaxies are stronger they obviously also impose limits on the total expected Galactic DM substructures contribution to the IGRB. Yet in these cases our limits are still relevant, as they represent an independent probe of cross sections by means of a conceptually different approach.

While the gamma-ray signal originating from Galactic substructure could appear reasonably isotropic, an important difference with the extragalactic signal is in the spectral shape: the extragalactic signal is redshifted and distorted by absorption on the EBL (c.f. eq.~(\ref{finaleq})) while the Galactic signal directly reflects the injection spectrum of gamma rays from DM annihilations and is generally harder. For that reason it is important to take this component properly into account, especially for heavy DM candidates, for which EBL absorption can severely limit high-energy gamma-ray intensities.

%% file: 3DMlimits.tex
\section{Constraints on WIMP signals} \label{sec:limits}

%%%%%%%%%%%%%%%%%%%%%%%%%%%%%%%%%%%%%%%%%%%%%%%%%%%%%%%%%%%%%%%%%%%%%%%%%%%%%%%%%%%%%%%%%%%%%%%%%%%%%%%%%%%%%%%%%%%%%%%%%%
\subsection{A review of the new \fermi LAT IGRB measurement}  \label{sec:reviewEGB}
Before we derive constraints on DM signals, let us summarize the four main steps that were taken in the analysis of ref.~\cite{EGBnew} to measure the IGRB, which will be used in this section and further discussed in the next section. In total, 50~months of LAT data were used, and dedicated cuts---creating two new event classes---were used to produce data samples with minimal contamination of CR backgrounds. In particular, the data were divided into a low-energy data set (events used to derive the spectrum in the 100 MeV to 12.8 GeV range) and a high-energy sample (12.8 GeV to 820 GeV). Stronger cuts were applied in the low-energy range, where the event statistics are better but CR contamination is higher%(P7REP\_IGRB\_LO event class)
; the cuts were loosened at high energies, where both event statistics and CR contamination are lower%(P7REP\_IGRB\_HI event class)
. The full-sky data were then analyzed as follows:
%%%%%%%%%%%%%%%%%%%%%%%%%%%%%%%%%%%%%%%%%%%%%%%%%%%%%%%%%%%%%%%%%%%%%%%%
\begin{enumerate}
\item
 The full sky gamma-ray emission was modeled with a series of templates. Templates of the Galactic diffuse emission are produced with the \galprop code\footnote{http://galprop.stanford.edu} \cite{FermiLAT:2012aa}, based on three distinct diffuse model setups, dubbed models A, B and C (see Appendix~\ref{appendix:limitsB} and \cite{EGBnew} for further details). The templates include maps of the gamma-ray emission due to interactions of CRs with interstellar gas and the inverse Compton (IC)  emission separately. In addition, templates modeling the IC emission of CR electrons in the Solar radiation field, diffuse emission from Loop I and point sources from the 2FGL catalog were used. After masking regions of bright interstellar emission along the Galactic plane, the normalization of each template was fitted individually in energy bins in the range between 100 MeV and 12.8 GeV.
\item
 Above 51.2 GeV the event statistics do not allow for fitting in individual energy bins. To handle low statistics at high energies, the low-energy data set in the range from 6.4 GeV to 51.2 GeV was used to find the best-fit normalizations of the Galactic templates.
 The normalizations were then fixed to these best-fit values, and \galprop's predictions of the spectral shapes were applied up to 820~GeV in order to perform IGRB fits above 12.8 GeV using the high-energy data sample.
\item
 The isotropic component thus derived is a sum of the IGRB emission and misclassified particle backgrounds. The IGRB is then obtained by subtracting a model for the CR contamination, obtained from Monte Carlo studies,\footnote{The Monte Carlo studies include a simulation of the relevant charged particle species and intensities present in the near-Earth environment as well as a phenomenological model for gamma-ray emission from the Earth limb.} from the isotropic emission.
\item
Using a baseline model for the Galactic diffuse emission (model~A in ref.~\cite{EGBnew}), the above procedure was then repeated for different values of the relevant CR parameters. The scatter among the different IGRB spectra derived in this way, together with those derived by assuming foreground models B and C in \cite{EGBnew}, represents the systematic error band, indicated in figure~12 of ref.~\cite{EGBnew}.
\end{enumerate}
%%%%%%%%%%%%%%%%%%%%%%%%%%%%%%%%%%%%%%%%%%%%%%%%%%%%%%%%%%%%%%%%%%%%%%%%
The result of this procedure is a measure of the spectrum of the IGRB in the range from 100 MeV to 820 GeV \cite{EGBnew}. It should be noted that this measurement is performed without including any Galactic smooth DM signal template, and the effects of such non-isotropic DM signal will be discussed in section~\ref{sec:gadgetchecks}.

%%%%%%%%%%%%%%%%%%%%%%%%%%%%%%%%%%%%%%%%%%%%%%%%%%%%%%%%%%%%%%%%%%%%%%%%%%%%%%%%%%%%%%%%%%%%%%%%%%%%%%%%%%%%%%%%%%%%%%%%%%
\subsection{Known source contributors to the IGRB}\label{sec:sources}

The \fermi LAT has detected many extragalactic sources: among the 1873 sources in the 2FGL catalog, there are 672 blazars (all classified according to the Roma BZCAT\footnote{v4.1, http://www.asdc.asi.it/bzcat/}), 8 radio galaxies, 3 normal galaxies, 3 starburst galaxies and 2 Seyfert galaxies \cite{2012ApJS..199...31N}.\footnote{There are 354 additional sources all associated in the 2FGL that appear to have blazar-like temporal or spectral characteristics but for which the lack of optical spectra did not allow a precise classification, most of them being labeled as AGN of uncertain type \cite{2011ApJ...743..171A}.}  The contribution to the IGRB from unresolved members of these extragalactic source classes has been studied over the years, \eg, \cite{2009ApJ...702..523I,2010ApJ...720..435A,Ajello:2013lka,DiMauro:2013zfa,Abazajian:2010pc,Stawarz:2005tq,Bhattacharya:2009yv,Inoue:2011bm,DiMauro:2013xta,Ackermann:2012vca,Calore:2013yia,Cholis:2013ena,Tamborra:2014xia}. In addition, some classes of Galactic sources, most notably millisecond pulsars, could contribute to the isotropic emission at large scale height in the Milky Way \cite{FaucherGiguere:2009df}. Their contribution to the IGRB is however severely constrained by the strong gamma-ray angular anisotropy signal expected for this source class \cite{SiegalGaskins:2010mp,Calore:2014oga}. There are other truly diffuse emission processes that are expected to contribute to the IGRB as well, although probably only at a few percent level, \eg  structure formation shocks in clusters of galaxies \cite{Zandanel:2014pva} and giant radio lobes of FR II radio galaxies \cite{2011ApJ...729L..12M}.  {We refer the reader to Ref.~\cite{Fornasa:2015qua} for a recent review on the IGRB and a discussion about potential contributors to this emission.}

Overall, the origin and composition of the IGRB are still open questions. Because of the large number of blazars detected by the %\fermi 
{LAT,} direct population studies are now feasible using gamma rays and  there is arguably a guaranteed contribution from the blazar population \cite{2009ApJ...702..523I,2010ApJ...720..435A,Ajello:2013lka,DiMauro:2013zfa,Abazajian:2010pc}. The {\it minimum} contribution below 100 GeV from unresolved blazars has been estimated in ref.~\cite{2010ApJ...720..435A} to be close to $10\%$, the best estimate being $22-34 \%$ in the 0.1-100 GeV range (which agrees well with previous findings, \eg \cite{2012ApJ...753...45S,2012JCAP...11..026H,DiMauro:2013zfa}). {The blazar contribution to the IGRB at the highest energies has only recently been studied. In ref.~\cite{DiMauro:2013zfa} they used a preliminary version of the new IGRB measurement reported in \cite{EGBnew} and concluded that blazars can naturally explain the total measured IGRB above 100 GeV. }

For the other known source classes, however, we lack this kind of direct information, and cross correlations with radio (in the case of radio galaxies, see, \eg,~\cite{Stawarz:2005tq,Bhattacharya:2009yv,Inoue:2011bm,DiMauro:2013xta}) or infrared data (for star-forming galaxies, \eg,~\cite{Ackermann:2012vca}) have been used to determine the luminosity functions and infer the expected intensities in the \fermi LAT energy range. 
In a companion \fermi LAT paper \cite{ajello14}, the contribution of blazars in the full energy range has been reevaluated using an updated luminosity function and spectral energy distribution model, taking advantage of recent follow-up observations \cite{Ajello:2013lka}. When summing the contribution from star-forming galaxies \cite{Ackermann:2012vca}, radio galaxies \cite{Inoue:2011bm} and blazars, ref.~\cite{ajello14} shows that these three contributors could account for the intensity of the EGB across the 0.1 - 820 GeV range sampled by \fermi LAT. In ref.~\cite{ajello14}, the methodology of this work was adopted to derive DM limits taking advantage of the aforementioned new estimates of the astrophysical contributions.
 {Yet, since %minimum 
intensity estimates for each of these potential IGRB contributors are uncertain (or under study) at the moment, in this work we stay agnostic about the precise contribution of the astrophysical populations to the IGRB and instead aim for a more general approach.}

%%%%%%%%%%%%%%%%%%%%%%%%%%%%%%%%%%%%%%%%%%%%%%%%%%%%%%%%%%%%%%%%%%%%%%%%%%%%%%%%%%%%%%%%%%%%%%%%%%%%%%%%%%%%%%%%%%%%%%%%%%
\subsection{Statistical analysis}  \label{sec:analysisapproach}

We form a test statistic ($TS$) with a presumed $\chi^2$ distribution: 
\be
TS = \chi^2 = \sum_{ij} (D_i-M_i) V_{ij}^{-1} (D_j-M_j), \label{eq:TS}
\ee
where $D_i$ is the measured IGRB intensity in energy bin $i$, $V^{-1}$ is the inverse of the variance-covariance matrix and $M_i$ is the IGRB model prediction (see below).

On top of statistical uncertainties, the IGRB measurement inherits significant systematic uncertainties from the effective-area and the CR-contamination determination.
These systematic uncertainties, combined with the IGRB measurement procedure (summarized in section~\ref{sec:reviewEGB}), can induce correlations between the IGRB measurements in the different energy bins. 
This should be reflected by a proper variance-covariance matrix $V$, and we made a study to estimate the variance-covariance matrix and check its impact on DM limits compared to the common approximation of taking it to be diagonal.

To establish the expected correlation matrix for the data, 1000 Monte Carlo-generated pseudo experiments were created. Gamma-ray sky maps were generated with the help of HEALPix\footnote{http://healpix.jpl.nasa.gov/} by taking the number of events in each sky pixel to be Poisson-distributed around the observed number of events.
The effective area and the CR contamination were drawn from normal distributions around their nominal values in \cite{EGBnew}. 
To account for bin-to-bin correlations of the systematic uncertainty of the effective area we included correlations on the scale of three adjacent energy bins \cite{2012ApJS..203....4A}. 
The CR contamination in the low- and high-energy data samples were taken to be fully uncorrelated. However, within each of these data samples, the systematic uncertainty (taken from \cite{EGBnew}) was used to induce a randomized overall shift factor for the CR contamination rate level. The remaining (subdominant) statistical uncertainties for the CR contamination were taken to be fully uncorrelated. 

Each Monte Carlo-generated data set was then used to perform IGRB measurements exactly as done with the real data (using model~A for the Galactic diffuse emission) and taking the CR and effective area determinations as described above. From the sample of 1000 IGRB pseudo measurements, the correlation matrix can then be directly calculated \cite{Cowan}. Subsequently, the variance-covariance matrix is determined by using the IGRB total variances (\ie the sum of the variances from statistical, CR contamination and effective area uncertainties) in each energy bin as they were given in ref.~\cite{EGBnew}.
The correlations are seen to be the strongest between neighboring energy bins at low energies, while energy bins further away and at the highest energies have negligible correlations. The derived variance-covariance matrix was then directly used in our statistical DM analysis (i.e.\ we included our calculated $V$ in the $TS$ calculation of eq.~(\ref{eq:TS})), where it typically induced about a factor two increase in $\chi^2$ but  less effects on $\Delta \chi^2$. 
From this study we concluded that the impact on DM limits from including a proper variance-covariance matrix can have a sizeable effect but are typically smaller than the shift in DM limits coming from changing the diffuse foreground modeling shown in Appendix~\ref{appendix:limitsB}.

For our final analysis, we therefore decided to treat all data points $D_i$ as uncorrelated with Gaussian probability distributions in our $\chi^2$ evaluations. The variance-covariance matrix $V$ is thus approximated as diagonal with elements ${\sigma_i}^2$. The systematic uncertainties of the effective area and the charged CR contamination as well as the statistical errors are added in quadrature and their sum is the ${\sigma_i}^2$ that enter in the covariance matrix $V$. The IGRB spectrum data points $D_i$ and the just specified 1-$\sigma$ errors $\sigma_i$ can be found in the supplementary tables of ref.~\cite{EGBnew}.\footnote{\url{https://www-glast.stanford.edu/pub_data/845/}}

In addition, there is a significant systematic uncertainty due to the assumed Galactic foreground emission model in the IGRB derivation. The investigated range of Galactic foreground emission assumptions can induce correlated IGRB data points, and uncertainties are presumably very asymmetric. Galactic foreground uncertainties will therefore not be included in the evaluation of the $\chi^2$ in eq.~(\ref{eq:TS}), but their impact was taken into account or estimated separately in our procedures, as we detail below.

%%%%%%%%%%%%%%%%%%%%%%%%%%%%%%%%%%%%%%%%%%%%%%%%%%%%%%%%%%%%%%%%%%
\subsubsection{WIMP signal search}\label{sec:search}
Before setting any limits, we will use a single power law with an exponential cut-off as a null-hypothesis background, 
  \be
  \frac{d\phi^\textrm{bkg}}{dE_0} = 
  A E_0^{-\alpha} \exp \left(-\frac{E_0}{E_c}\right),
  \label{eq:pow}
  \eeq
and search for any significant detection if a WIMP signal is added on top of this. This background should be viewed 
as the effective IGRB spectrum from all conventional astrophysical sources discussed in section~\ref{sec:sources}, where the overall normalization $A$, spectral index $\alpha$ and exponential cut-off energy $E_c$ are free parameters. 

For the DM signal search we use the IGRB derived with model~A for the Galactic diffuse emission and only {the $\sigma_i$ errors} are included in the $\chi^2$ calculations. The simple background model of eq.~(\ref{eq:pow}) gives an excellent best-fit to the 26 data points of the IGRB spectrum; with a  $\chi^2$ of 13.7  for 23 degrees-of-freedom.\footnote{This small $\chi^2$ value is presumably related to the fact that {LAT's systematic effective area and CR contamination uncertainties are included in the $\sigma_i$ while their induced correlations are ignored in the variance-covariance matrix. If we include the variance-covariance matrix discussed in section~\ref{sec:analysisapproach} then $\chi^2$ become 21.0; which is in good agreement with what should be expected.}} Naively this leaves little need for including an additional DM signal. 
To still search for a statistically significant DM signal we use the DM set-up described in section~\ref{2Theory}, where the isotropic DM signal is the sum of the contributions from cosmological signal and Galactic substructures. For the cosmological calculation we use the HM result, shown by the red solid line in figure~\ref{fig:PS_HM_comp}. Also, since the ratio of DM extragalactic to Galactic substructure signals affects the total DM signal spectral shape, we investigate the theoretical uncertainty range in the extragalactic signal strength by using the lowest and highest results from the PS approach (given by the limits of the gray band in figure~\ref{fig:PS_HM_comp}). As for the Galactic substructures, in combination with either the HM or PS cosmological signals, we use two different overall signal strengths (corresponding to boosts of 3 and 15 for the Galaxy DM halo signal as a whole). The different WIMP annihilation channels we test are the same as those specified in section~\ref{sec:wimpconstraints}, where we present our WIMP annihilation cross-section limits and derive our sensitivity reach.

The $\chi^2$ difference between the best-fit including such additional WIMP signal on top of the background (with the WIMP annihilation cross section as one extra degree of freedom) and the null hypothesis with zero DM signal, reveals that none of the DM hypotheses we tested showed a fit improvement by more than  $\Delta TS  = 8.3$. Assuming that our $TS$ has a $\chi^2$ distribution, the local significance is 2.9$\sigma$ (before including any trials factor). The largest significance was in our minimal setup for both the extragalactic signal and the Galactic substructure signal, with a WIMP mass of 500~GeV and an annihilation cross section into $\mu^+\mu^-$ pairs of $1.1\times 10^{-23}$ cm$^3$/s. Note that this significance is not large, especially since a trials factor has not been yet applied (among 192 models we tested we had more than $2\sigma$ detections for 16 different models).
More importantly, uncertainties in the IGRB related to the selected Galactic diffuse emission model were not included. In fact, we also performed the analysis with IGRB derived with Galactic diffuse model~B and C and confirmed the DM non-detection obtained with model~A. For example, the $\Delta TS = 8.3$ mentioned above drops to 3.9 when the IGRB derived with the Galactic diffuse model B is used.

We test at most nine WIMP masses for each annihilation channel: 10, 50, 100, 500, 1000, 5000, 10000, 20000, and 30000 GeV.  This leaves a possibility that a somewhat more significant detection could be found if smaller steps in WIMP mass were used.

For the tested WIMP masses we therefore conclude that there is no clear statistically significant evidence of WIMP signals in the IGRB, and we proceed to calculate upper limits. This is a rather naive approach and, indeed, a better understanding of the astrophysical contributions to the IGRB could help in revealing potential anomalies or even point toward the need for a DM contribution to the measured IGRB.

%%%%%%%%%%%%%%%%%%%%%%%%%%%%%%%%%%%%%%%%%%%%%%%%%%%%%%%%%%%%%%%%%%
\subsubsection{{Conservative approach for setting WIMP limits and the sensitivity reach of the IGRB measurement}} \label{sec:ConOpt_approach}

We will focus on deriving i) {\it conservative DM limits}, derived by making assumptions {\emph{neither} on the contributions from unresolved astrophysical populations to the IGRB nor on a specific choice of a Galactic diffuse emission model,} and {ii) {\it sensitivity reach}, which assumes both that the total contribution from conventional astrophysical sources fully explains the measured IGRB at all energies by eq.~(\ref{eq:pow}), and that we can entirely rely on a specific Galactic diffuse foreground model to derive the IGRB.

We adopt the following two procedures: 
\begin{itemize} 
\item
{\bf Conservative  limits:} The $\chi^2$ in eq.~(\ref{eq:TS}) is calculated \emph{only} for bins where the DM signal alone exceeds the {IGRB intensity}:
\be
 \chi^2_\textrm{cons} = \sum_{i \in \{ i | \phi_i^\mathrm{DM}>D_i^\mathrm{max}\} } \frac{\left[ D_i^\mathrm{max} - \phi^\mathrm{DM}_i(\langle \sigma v\rangle) \right]^2}{\sigma_i^2},
 \label{eq:chi2cons}
\ee
where $\phi^\mathrm{DM}_i$ is the integrated DM-induced intensities in energy bin $i$ as a function of $\langle \sigma v\rangle$ for a given WIMP candidate. Effectively, this corresponds to having a (non-negative) background model that is free in normalization in each energy bin independently.  {To have a rough estimate of the effect from having various Galactic foreground emission models that can alter the IGRB,} we shift the centers of the IGRB data points derived with Galactic emission model~A (while we keep the size of the errors $\sigma_i$) to the upper edge of the 1-$\sigma$ envelope of all tested Galactic diffuse models used in ref.~\cite{EGBnew}.\footnote{ {We note that, in principle, one should marginalize over all the possible Galactic foreground models. Given that this is impractical, we opted here for a simpler approach that we believe provides a reasonable estimate of the effect.}}
This yields our data points $D_i^\mathrm{max}$. The mentioned shift of the IGRB bins reflects the conservative approach of considering all tested diffuse emission models in ref.~\cite{EGBnew} equally likely for determining the IGRB (maximal) intensity in every energy bin. The 2$\sigma$ (3$\sigma$) DM limits are then defined to be the cases in which the DM signal component gives $\chi^2_\textrm{cons}$ equal to 4 (9).\footnote{To avoid potential issues with correlations among the data points, we also performed the exercise of setting our limits by using {\it one single bin} (the one in which DM signal to the IGRB flux ratio is maximal) instead of a sum over bins as in eq.~(\ref{eq:chi2cons}). We found that the two approaches lead to limits which differ by only $\lta ~10\%$.} 

\item

{ {\bf Sensitivity reach:}} In this procedure, we use the IGRB derived with Galactic diffuse model~A and include the DM signal on top of the background model in eq.~(\ref{eq:pow}). The $\chi^2$ of eq.~(\ref{eq:TS}) is then evaluated over all energy bins:
\be
 \chi^2_\textrm{sens} = \sum_{i} \frac{\left[D_i - \phi^\mathrm{bkg}_i(A,\alpha,E_c)  - \phi^\mathrm{DM}_i(\langle \sigma v\rangle) \right]^2}{\sigma_i^2},
\ee

where $\phi^\mathrm{DM}_i$ + $\phi^\mathrm{bkg}_i$ is the model prediction $M_i$ {as a function of $\langle \sigma v\rangle$ for a given WIMP candidate. Note that this represents a scenario in which i) the Galactic diffuse foreground used to derive the IGRB is fixed, ii) the contribution from conventional astrophysical sources to the IGRB is described by the parametric form of eq.~(\ref{eq:pow}) and iii) the parameters $A$, $\alpha$, $E_c$ in eq.~(\ref{eq:pow}) are {\it fixed} to their best-fit values found in ref.~\cite{EGBnew} (given in their table 4).} The 2$\sigma$ (3$\sigma$) limits are then defined to be the cases in which the DM signal component forces the $\chi^2_\textrm{sens}$ to increase by more than 4 (9)  with respect to the  best-fit $\chi^2_\textrm{sens}$ with a free DM signal normalization. 
\end{itemize}

We have also checked how {this sensitivity reach changes} by varying the adopted Galactic foreground model, namely by comparing limits when the IGRB is derived with the alternative foreground models A, B and C in ref.~\cite{EGBnew}. {With this exercise, we gauge the impact of some systematic uncertainties associated with the modeling of the Galactic diffuse emission.} We find differences that can be substantial especially for low WIMP masses; see Appendix~\ref{appendix:limitsB} for further details. Yet, it should be noted that these tests are far from comprehensive and, as such, might not address the full range of uncertainties.
%%%%%%%%%%%%%%%%%%%%%%%%%%%%%%%%%%%%%%%%%%%%%%%%%%%%%%%%%%%%%%%%%%%%%%%%

{The sensitivity reach derived here can also be taken as limits under the given assumptions. However, strictly speaking they should be interpreted as DM constraints only if the astrophysical background was independently  predicted to the spectrum of eq.~(\ref{eq:pow}) with parameters equal to the best-fit values from the current IGRB measurement.}  

{The case where the total contribution to the IGRB from conventional astrophysics is derived as accurately as possible leads to DM constraints that typically lie between the conservative limit and the sensitivity reach derived in this work.} Indeed, this is what is obtained in a companion work \cite{ajello14}, where unresolved astrophysical source populations were modeled and used to set new DM limits on DM annihilation cross sections.

In figures~\ref{fig:CONSlimits_fluxes} and \ref{fig:OPTlimits_fluxes} we show illustrative examples of DM-induced spectra which have DM annihilation cross sections at the size of our 95\% CL exclusion limits by our conservative approach and our sensitivity reaches, respectively.
%
%%%%%%%%%%%%%%%%%%%%%%%%%%%%%%%%%%%%%%%%%%%%%%%%%%%%%%%%%%%%%%%%%%%%%%%%
\begin{figure}[t!]
\centering
\includegraphics[width=0.99\textwidth]{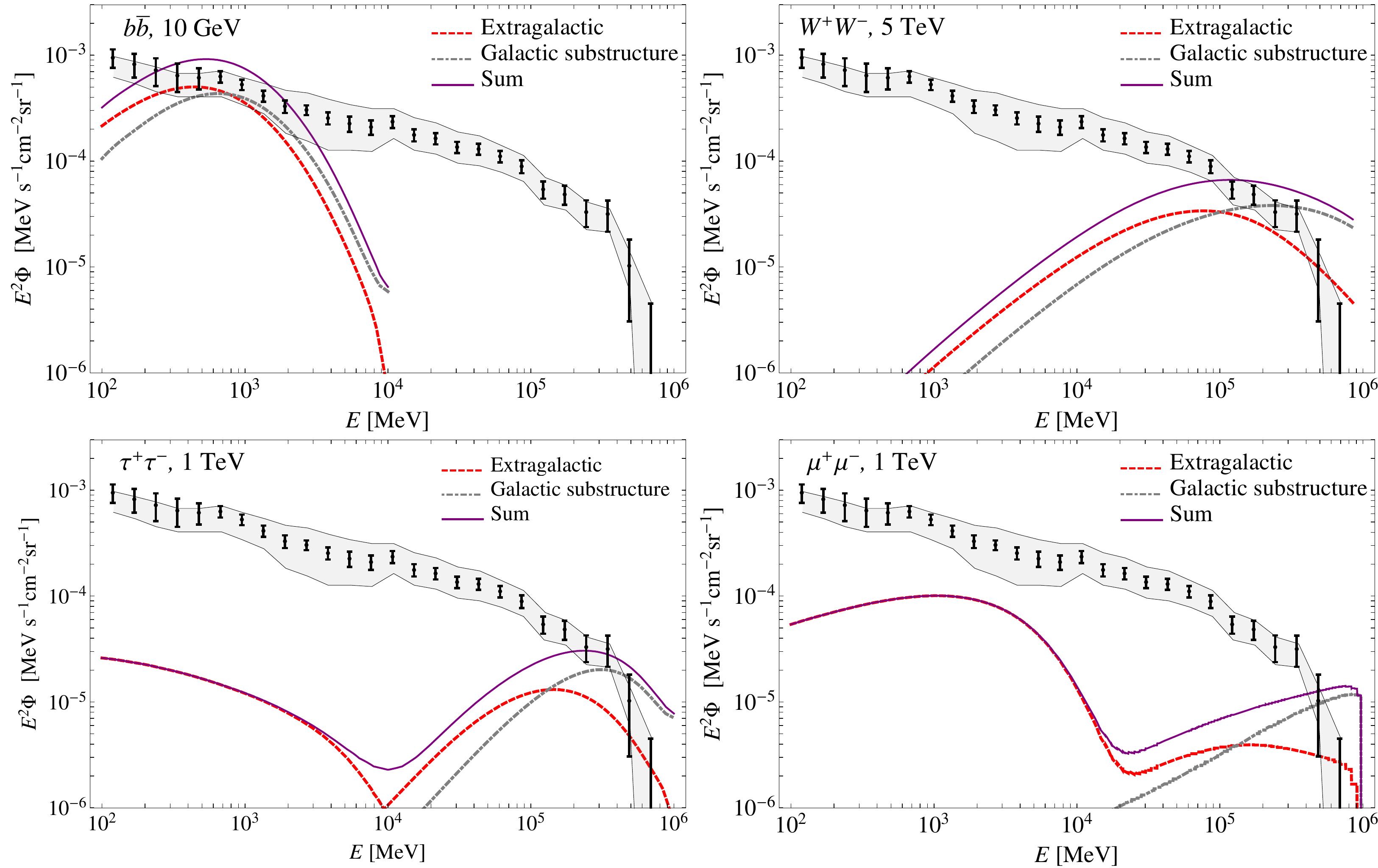}
\caption{\small{{Examples of DM-produced gamma-ray spectra which are at the border of being excluded by our 2$\sigma$ conservative limits. The WIMP mass and its annihilation channel is given in the upper left corner of each panel. The normalizations of the extragalactic signal and of the Galactic substructure signal are given by our benchmark HM model, as defined in section~\ref{subsec:HM}. Data points are in black, and the black lines show the upper and lower envelopes of the systematic uncertainties defined as the scatter among the different IGRB spectra derived in ref.~\cite{EGBnew}.}}}
\label{fig:CONSlimits_fluxes}
\end{figure}
%%%%%%%%%%%%%%%%%%%%%%%%%%%%%%%%%%%%%%%%%%%%%%%%%%%%%%%%%%%%%%%%%%%%%%%%
%

%
%%%%%%%%%%%%%%%%%%%%%%%%%%%%%%%%%%%%%%%%%%%%%%%%%%%%%%%%%%%%%%%%%%%%%%%%
\begin{figure}[t!]
\centering
\includegraphics[width=0.99\textwidth]{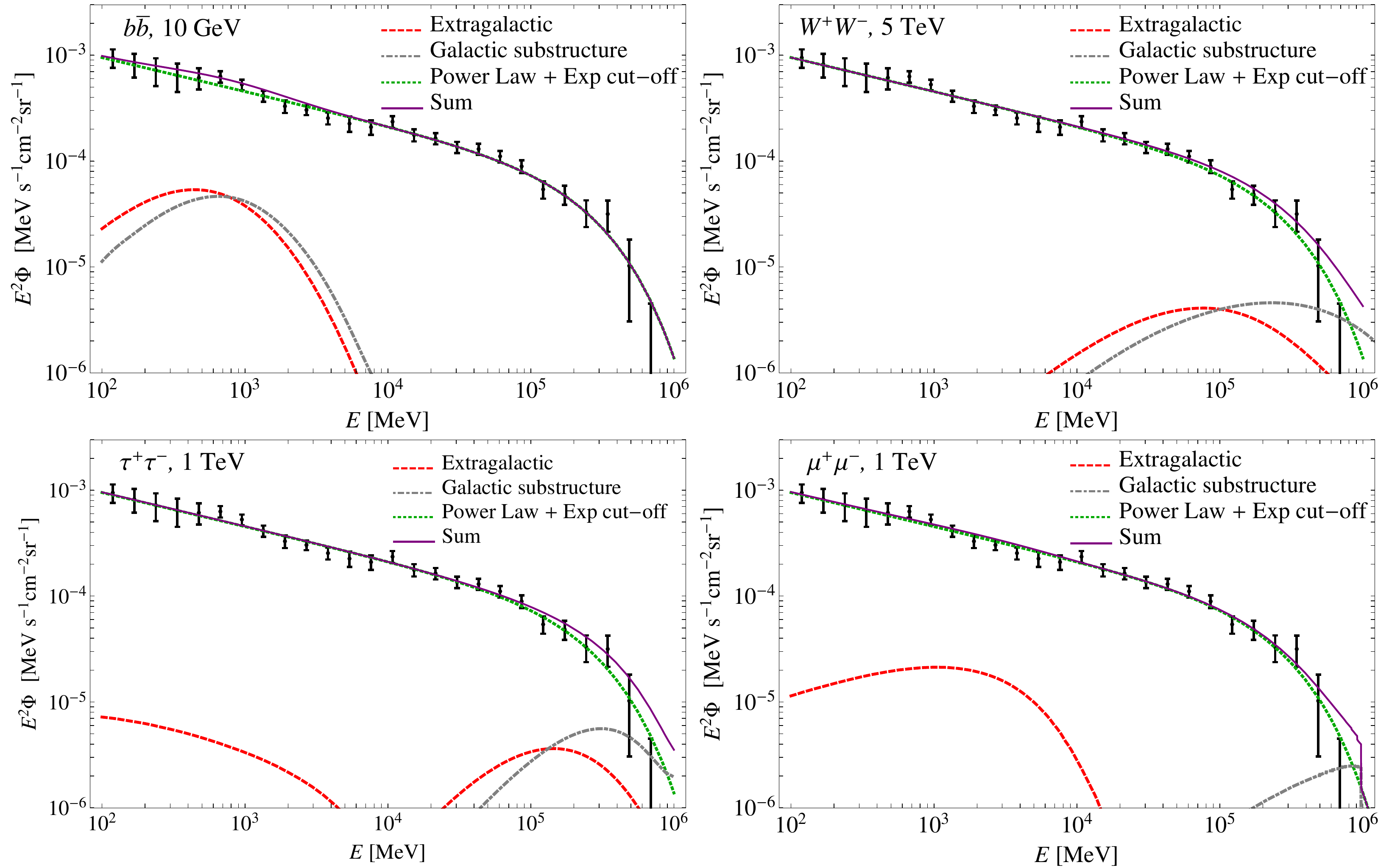}
\caption{\small{{Examples of DM-produced gamma-ray spectra which are at the border of being excluded at 2$\sigma$ level in our procedure to calculate the sensitivity reach of the IGRB data. The WIMP mass and its annihilation channel is given in the upper left corner of each panel. The normalizations of the extragalactic signal and of the Galactic substructure signal are given by our benchmark HM model, as defined in section~\ref{subsec:HM}. Data points from ref.~\cite{EGBnew}.}}}
\label{fig:OPTlimits_fluxes}
\end{figure}
%%%%%%%%%%%%%%%%%%%%%%%%%%%%%%%%%%%%%%%%%%%%%%%%%%%%%%%%%%%%%%%%%%%%%%%%
%

%%%%%%%%%%%%%%%%%%%%%%%%%%%%%%%%%%%%%%%%%%%%%%%%%%%%%%%%%%%%%%%%%%%%%%%%%%%%%%%%%%%%%%%%%%%%%%%%%%%%%%%%%%%%%%%%%%%%%%%%%%
\subsection{Limits on WIMP annihilation cross sections}\label{sec:wimpconstraints}
In this work, we stay agnostic about the nature of the DM particle and consider generic models in which DM annihilates with 100\% branching ratio into $b{\bar b}$, $W^+W^-$, $\tau ^+ \tau ^-$ or $\mu^+\mu^-$ channels. 
 {For the first two channels, we consider only prompt emission and do not include any secondary gamma rays coming from the DM-induced electrons that up-scatter CMB photons by IC. Even for the heaviest DM masses we consider, the prompt emission is soft enough here to contribute significantly within the energy range of the measured IGRB -- while the IC can be ignored because it only induces emission at much lower energies where the IGRB flux is higher. For $\tau^+\tau^-$ and $\mu^+\mu^-$ channels, instead, the prompt emission is harder and peaks significantly above the energy range for which the IGRB has been measured for our highest DM masses. In these cases the IC (which is also harder than for the previous two channels) contributes significantly at energies close to the observed IGRB exponential cut-off and thus must be included.
%\red{For the first two channels we consider the prompt emission only, because this emission is softer than for the $\tau ^+ \tau ^-$ or $\mu^+\mu^-$ channels, and even for the heaviest DM masses we consider, it still contributes significantly in the energy range before the IGRB cuts off. In addition the electron spectra produced in this case is very soft and the corresponding IC emission contributes only at low energies, where IGRB fluxes are high. For $\tau ^+ \tau ^-$ or $\mu^+\mu^-$ channels, instead, the prompt emission is harder and drops significantly in the IGRB energy range for high masses we consider. In that case the IC (which is harder than for the previous two channels) contributes significantly close to the IGRB exponential cut-off.} 
For that reason, both annihilation channels prove} to be especially strongly constrained by the IGRB measurement \cite{Abdo:2010dk}. We calculate the DM annihilation prompt spectra using the publicly available \texttt{PPPC4DMID} code \cite{Cirelli:2010xx}, which takes into account electroweak bremsstrahlung corrections, which are particularly relevant for heavy DM candidates. For the calculation of the IC emission from the muon channel we follow the calculation presented in \cite{Abdo:2010dk}.

%For the first two channels we consider the prompt emission only, because for these channels generally less energy goes to electrons with respect to the production of gamma rays, and secondary emission from IC electron scattering on cosmic microwave background (CMB) photons is expected to be subdominant. On the other hand, annihilation to muons induces a hard gamma-ray spectrum due to the Final State Radiation (FSR) emission and a large amount of hard electrons results in a pronounced intensity enhancement at lower energies due to the electron scattering on the CMB, which is present with increased intensity at higher redshifts. For that reason this annihilation channel proves to be especially strongly constrained by the IGRB measurement \cite{Abdo:2010dk}. We calculate the DM annihilation prompt spectra using the publicly available \texttt{PPPC4DMID} code \cite{Cirelli:2010xx}, which takes into account electroweak bremsstrahlung corrections, which are particularly relevant for heavy DM candidates. For the calculation of the IC emission from the muon channel we follow the calculation presented in \cite{Abdo:2010dk}. 

{For the four annihilation channels under consideration, we present the conservative limits and cross-section sensitivity reach at the 2$\sigma$ confidence level in figures~\ref{fig:CONSlimits_sigv} and \ref{fig:OPTlimits_sigv}, respectively.} In all cases, the DM limits were obtained by adopting the cosmological DM annihilation induced gamma-ray intensities given by the HM setup described in section~\ref{subsec:HM}, as well as a theoretical uncertainty range as estimated within the PS approach of section~\ref{subsec:PS} (gray band in figure~\ref{fig:PS_HM_comp}). In addition, {two configurations for the Galactic substructure contribution}---which is assumed to be isotropic in this work---are adopted: i) the \emph{reference} case, {labeled as "SS-REF" in figures~\ref{fig:CONSlimits_sigv} and \ref{fig:OPTlimits_sigv}}, where substructures boost the total Galactic annihilation signal by a factor of 15, and ii) the \emph{minimal} case, labeled "SS-MIN" in the figures, where the boost from Galactic substructure is 3. {Conservative DM limits and cross-section sensitivities} at the 3$\sigma$ level for the $b{\bar b}$ and $\tau ^+ \tau ^-$ channels were also derived, and can be found in Appendix~\ref{appendix:3sigma}. 
% 
%%%%%%%%%%%%%%%%%%%%%%%%%%%%%%%%%%%%%%%%%%%%%%%%%%%%%%%%%%%%%%%%%%%%%%%%
\begin{figure}[th!]
\centering
\includegraphics[width=0.99\textwidth]
{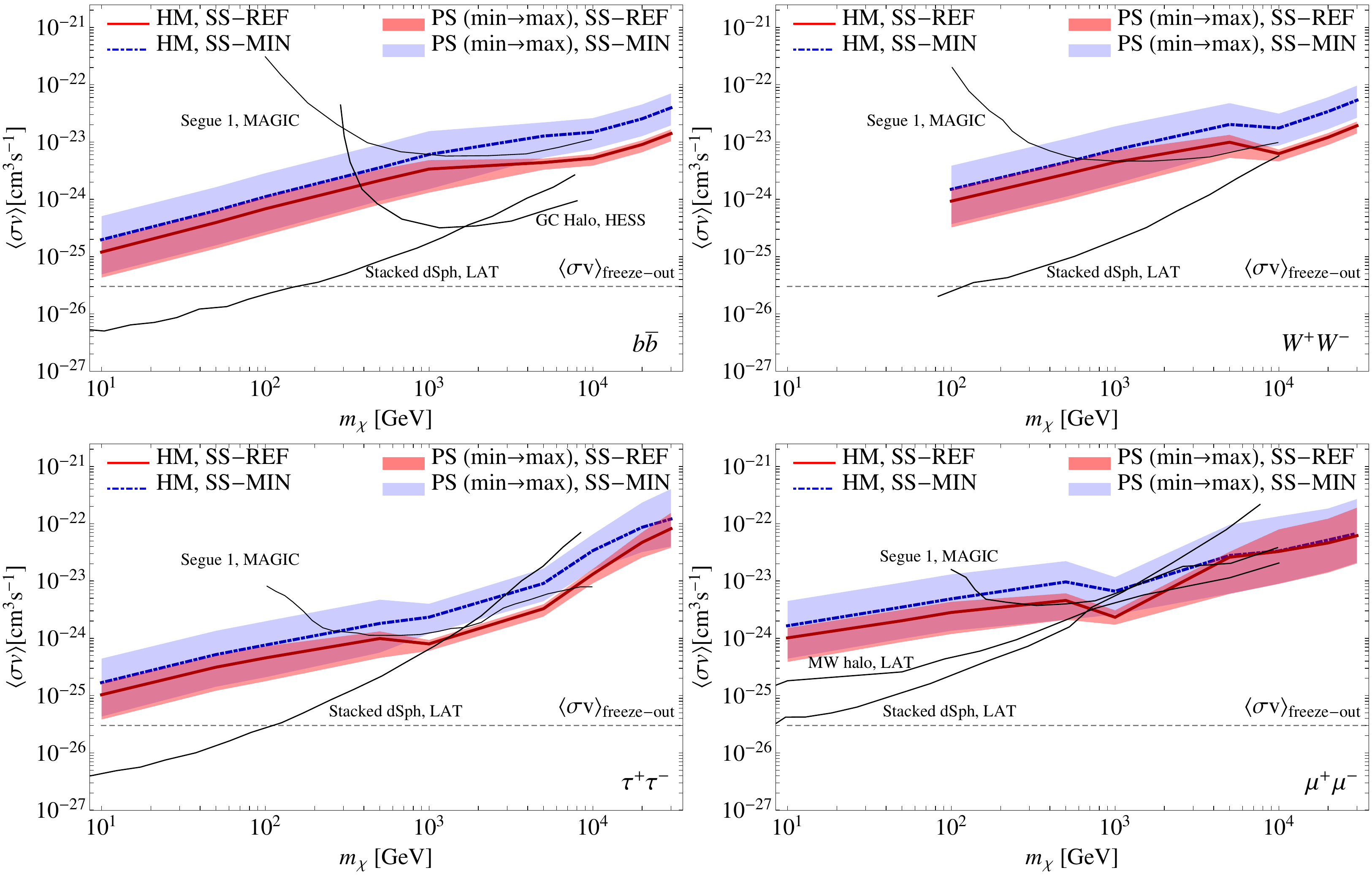}
\caption{\small{{Upper limits (95\% CL) on the DM annihilation cross section in our conservative procedure. From top to bottom and left to right, the limits are for the $b{\bar b}$, $W^+W^-$, $\tau ^+ \tau ^-$ and $\mu^+\mu^-$ channels. The {\it red} solid line shows limits obtained in our {\it fiducial} HM scenario described in section~\ref{subsec:HM}, and assumes the reference contribution from the Galactic subhalo population; see section~\ref{sec:GalacticDM} (`HM, SS-REF' case). The broad red {\it band} labeled as `PS (min$\rightarrow$max), SS-REF' shows the theoretical uncertainty in the extragalactic signal as given by the PS approach of section~\ref{subsec:PS}. The {\it blue} dashed line (`HM, SS-MIN') , with its corresponding uncertainty band (`PS (min$\rightarrow$max), SS-MIN'), refers instead to the limits obtained when the Milky Way substructure signal strength is taken to its lowest value as calculated in ref.~\cite{2014MNRAS.442.2271S}.  For comparison, we also include other limits derived from observations with \fermi LAT \cite{2015arXiv150302641F,Ackermann:2012rg} and imaging air Cherenkov telescopes \cite{Abramowski:2011hc,Aleksic:2013xea}.}}}
\label{fig:CONSlimits_sigv}
\end{figure}
%%%%%%%%%%%%%%%%%%%%%%%%%%%%%%%%%%%%%%%%%%%%%%%%%%%%%%%%%%%%%%%%%%%%%%%%
%

%%%%%%%%%%%%%%%%%%%%%%%%%%%%%%%%%%%%%%%%%%%%%%%%%%%%%%%%%%%%%%%%%%%%%%%%
\begin{figure}[th!]
\centering
\includegraphics[width=0.99\textwidth]
{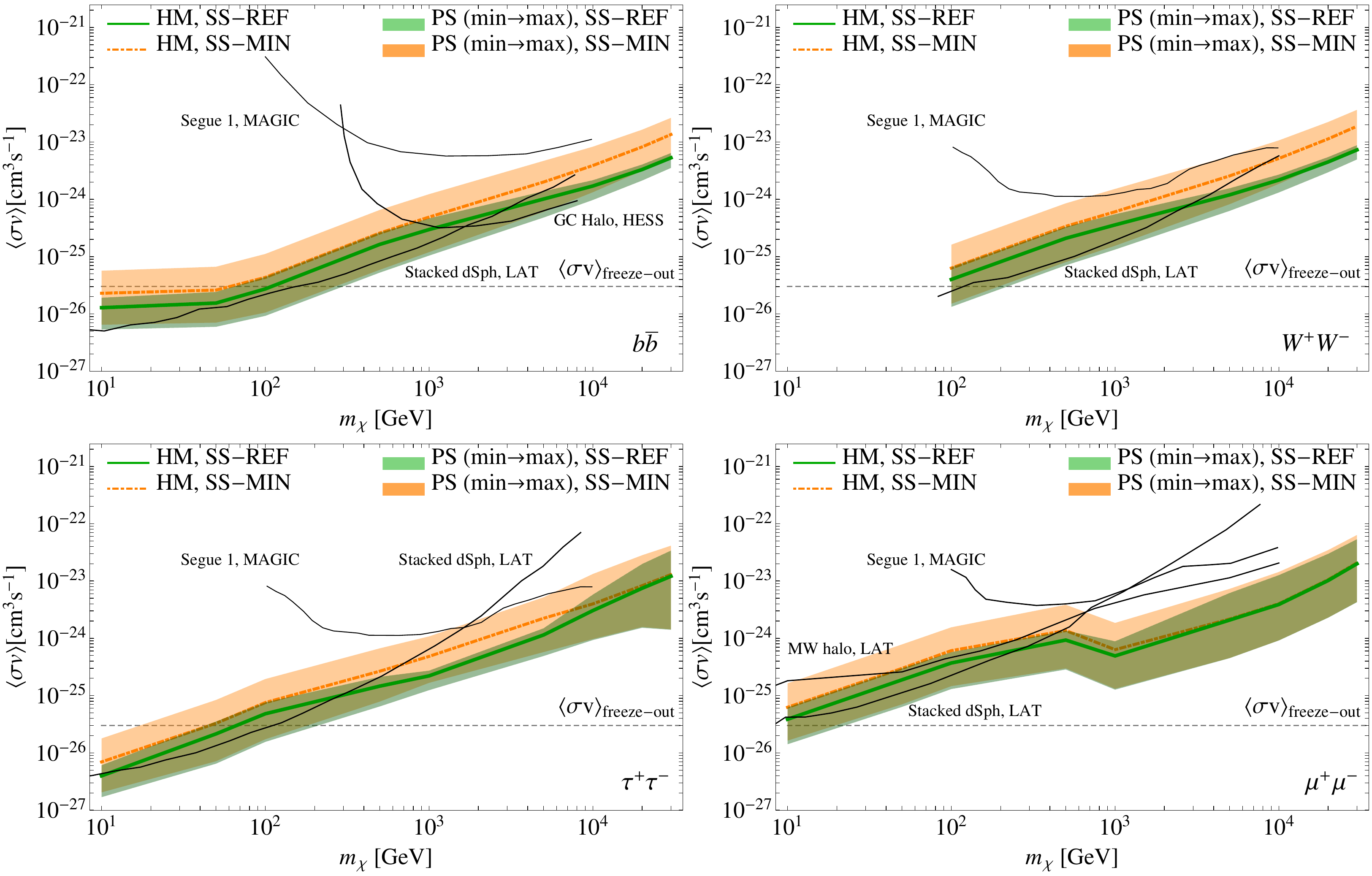}
\caption{\small{{DM annihilation cross section sensitivity reach (95\% CL). {\it Green} solid line shows sensitivity obtained in our {\it fiducial} HM scenario described in section~\ref{subsec:HM}, and assumes the reference contribution from the Galactic subhalo population; see section~\ref{sec:GalacticDM} (`HM, SS-REF' case in the panels). The broad green {\it band} labeled as `PS (min$\rightarrow$max), SS-REF' shows the theoretical uncertainty in the extragalactic signal as given by the PS approach of section~\ref{subsec:PS}. The {\it orange} dashed line (`HM, SS-MIN'), with its corresponding uncertainty band (`PS (min$\rightarrow$max), SS-MIN'), refers instead to the cross-section sensitivity obtained when the Milky Way substructure signal strength is taken to its lowest value as calculated in ref.~\cite{2014MNRAS.442.2271S}.  For comparison, we also include other limits derived from observations with \fermi LAT \cite{2015arXiv150302641F,Ackermann:2012rg} and imaging air Cherenkov telescopes \cite{Abramowski:2011hc,Aleksic:2013xea}.}}}
\label{fig:OPTlimits_sigv}
\end{figure}
%%%%%%%%%%%%%%%%%%%%%%%%%%%%%%%%%%%%%%%%%%%%%%%%%%%%%%%%%%%%%%%%%%%%%%%%
%

From theoretical considerations, various DM particle candidate masses span a huge range. For thermally produced WIMPs, however, the Lee-Weinberg limit restricts the mass to be above few GeV \cite{Lee:1977ua} and unitarity considerations bound it to be below $\sim 100$ TeV \cite{Griest:1989wd}. Interestingly, we are able to constrain signals for WIMP masses up to $\sim 30 $ TeV because the IGRB measurement now extends up to 820 GeV. For DM particle masses above $\sim 30 $ TeV, we start to probe the low-energy tail of the DM spectra and thus we lose constraining power rapidly. Furthermore, extragalactic WIMP signals are heavily suppressed at the highest energies as the optical depth is very large for such gamma rays. 

{It is interesting to compare the conservative limits of figure~\ref{fig:CONSlimits_sigv} to the cross-section sensitivities in figure~\ref{fig:OPTlimits_sigv}, at least for the case of our fiducial HM scenario and the reference contribution from the Galactic subhalo population (`HM, SS-REF' case in the panels). For  the $b\bar{b}$ ($\tau^+\tau^-$) channel, the differences are of about factors 9, 25, 11, 3 (26, 9, 4, 3) at 10 GeV, 100 GeV, 1 TeV, 10 TeV.}

{For low WIMP masses, the full spectral shape of the IGRB is affected by the WIMP signal, and hence the sensitivity reach, assuming a known spectral shape for the astrophysical contributions to the IGRB, places stronger limits, whereas for the largest WIMP masses only the last point(s) in the IGRB spectrum is affected and the two approaches are more similar.}\footnote{{If we omit the last data point, we find that both conservative limits and cross-section sensitivity for the $b{\bar b}$ channel worsen by $\lsi$30\% at 5 TeV mass going up to a factor of $\sim 2$ for masses between 10 and 30 TeV. In the case of the harder $\tau^+\tau^-$ channel, limits and sensitivity reach progressively weaken by a factor $\sim 2$ to 4 between 2 and 30 TeV, respectively.}}

For the largest WIMP masses considered, {the signal from Galactic substructures is stronger than that from the extragalactic DM}, with the effect that the uncertainty range of the extragalactic WIMP signal becomes irrelevant {when setting DM limits and calculating cross-section sensitivities.} This is typically the case for gamma-ray energies above 100~GeV, where extragalactic signals are effectively attenuated due to EBL attenuation. The effect can be clearly seen in figures~\ref{fig:CONSlimits_fluxes} and \ref{fig:OPTlimits_fluxes} for several annihilation channels (see, \eg, the spectra of a 5 TeV mass DM particle annihilating to the $W^+W^-$). As a result, as the WIMP mass increases in figures~\ref{fig:CONSlimits_sigv} and \ref{fig:OPTlimits_sigv}, the cross-section limit uncertainties get narrower (for a given Galactic substructure signal strength). For the same reason, the uncertainty band for the minimal Galactic substructure scenario ('SS-MIN' case in figures~\ref{fig:CONSlimits_sigv} and \ref{fig:OPTlimits_sigv}) is typically wider than the one for the reference Galactic substructure case (`SS-REF'), especially at the largest WIMP masses considered. {This is less pronounced for the muon channel, because in that case the high-mass limits are still set by the IC peak of the emission which contributes at low energies.}

Another feature worth mentioning is that, in the case of DM annihilation into $\mu ^+ \mu^-$, figures~\ref{fig:CONSlimits_sigv} and \ref{fig:OPTlimits_sigv} show a dip in cross-section limits for DM particle masses around 1~TeV. This dip is present because the part of the gamma-ray spectrum induced by FSR peaks at energies where the IGRB intensity has dropped exponentially above a few hundred GeV (see figure~\ref{fig:CONSlimits_fluxes}). For larger WIMP masses, the FSR peak is well above the energy range covered by the %\fermi 
{LAT} IGRB measurement and, as the WIMP mass increases, the limits get progressively weaker until lower-energy gamma rays---induced by IC upscattering of CMB photons from DM-induced high-energy electrons---eventually govern the constraints.

Finally, it is also interesting to compare the {conservative limits and cross-section sensitivities obtained in this work to other DM limits} recently reported in the literature. At low masses, $\lsi 100$ GeV, {the derived sensitivity reach is} comparable to the limits derived from a stacking analysis of 25 dwarf spheroidal satellites of the Milky Way \cite{2015arXiv150302641F}, and to the limits derived from considering diffuse emission at intermediate Galactic latitudes \cite{Ackermann:2012rg}. We note that the present analysis uses {LAT data} up to very high energies (820 GeV), which represents a novelty with respect to previous \fermi LAT DM searches.  In order to put in perspective the {results} derived here, we also compare to DM constraints derived from ground-based atmospheric Cherenkov telescope observations. {Figures~\ref{fig:CONSlimits_sigv} and \ref{fig:OPTlimits_sigv} show the limits derived by the \hess Collaboration from observing the Galactic center halo \cite{Abramowski:2011hc} and the \magic Collaboration's limits derived from deep observations of the Segue~1 dwarf spheroidal galaxy \cite{Aleksic:2013xea}.}
{Our conservative limits are comparable to the latter ones in the TeV energy range, but weaker than those obtained by \hess\footnote{We caution the reader that the \hess limits are derived under the assumption of a cuspy profile, whereas DM limits from IGRB data are only moderately sensitive to the inner slope of the DM halo density profiles.} As for the cross-section sensitivity reach, this is substantially better than the mentioned \magic limits, and comparable to the ones from \hess} Our work, which uses the IGRB's {\it total intensity} to set constraints on the nature of DM, is connected to studies focused on small-scale angular {\it anisotropies} in the high-latitude gamma-ray sky \cite{2013MNRAS.429.1529F,Ando:2013ff,Gomez-Vargas:2013cna}. Indeed, they both test for the presence of the same gamma-ray source class. In this regard the limits are directly comparable, and their DM signal limits are currently of similar magnitude \cite{Ando:2013ff,Gomez-Vargas:2013cna}. It has also been shown that a cross-correlation of an anisotropic signal with the positions of galaxies in the 2MASS survey \cite{Ando:2013xwa} or with weak lensing surveys \cite{Camera:2012cj,Camera:2014rja} could increase the ability to detect a DM component.

%% file: 4GadgetchecksSHORT.tex
\section{Robustness of the IGRB measurement in the presence of a Galactic dark matter signal component}\label{sec:gadgetchecks}

The largest component of the systematic uncertainty in the measurement of the IGRB spectrum stems from the modeling of the Galactic diffuse emission \cite{EGBnew}. A signal from the smooth Milky Way DM component would contribute to Galactic diffuse emission and, as seen in ref.~\cite{Ackermann:2012rg}, this signal can in part be degenerate with conventional astrophysical emissions. Indeed, the DM component is morphologically similar to the inverse Compton astrophysical emission and, in ref.~\cite{EGBnew}, it is seen that uncertainties in the IC template have the most significant impact on the measured IGRB spectrum. 

In order to study this issue we repeat the fitting procedure in \cite{EGBnew} and model the Galactic diffuse emission, but with the addition of a Galactic smooth DM template. Our aim is twofold: i) to verify that DM and Galactic diffuse templates are {partially degenerate with each other} in the fitting procedure, and ii) to {\it a posteriori} check {for} self-consistency of our procedure, \ie for those DM annihilation cross sections constrained in section~\ref{sec:wimpconstraints}, we test whether the corresponding Galactic DM counterpart emission alters the IGRB measurement that was used to set the DM limits themselves.

{Following ref.~\cite{EGBnew}, we perform low- and high-energy template fits separately (as described in section~\ref{sec:reviewEGB}). We use templates of  10, 50, 250 and 500 GeV, and 1, 5, 20 and 30 TeV DM particle masses annihilating to $b{\bar b}$ quarks and $\tau^+ \tau^-$ leptons}, as representative gamma-ray signals from WIMP annihilation.\footnote{Note that the morphology of the Galactic DM templates is independent of the DM annihilation channel considered, as long as one refers to the photons induced `directly' in the annihilation process.} We produce full-sky templates using the \galprop code version 54 \cite{2007ARNPS..57..285S} into which we have incorporated the DM signal. 
We then evaluate the maximum value of the DM annihilation cross section (for each DM mass) which still leaves the IGRB measurement unchanged. 
The IGRB is taken as `unchanged' when, after the inclusion of the Galactic smooth DM template, the {\it new} IGRB measurement for all energy bins falls inside either i)  
twice the width of the systematic uncertainty band derived from foreground model variations or ii) twice the 1$\sigma$ `statistical' error bars} of the IGRB measurement of ref.~\cite{EGBnew}. This maximal normalization can be translated into a maximal cross section once a particular Galactic DM density distribution is adopted. To be conservative---in the sense of finding the corresponding DM annihilation cross-section values in our Galaxy which are `guaranteed' to modify the IGRB---we set the local DM density to a low value of $\rho_0=0.2$ GeV~cm$^{-3}$ while keeping the Milky Way scale radius at the standard value of $r_s=20$ kpc, see \eg~\cite{Cirelli:2010xx}.\footnote{Strictly speaking, $r_s$ and $\rho_0$ are not independent, but for our purposes lowering $\rho_0$ alone serves as a reasonable way to adopt a low but realistic Galactic DM signal.}

Interestingly, while repeating the above procedure with different normalizations of the Galactic smooth DM templates we found that, for DM masses {below $\sim 1$ TeV}, an increase of the Galactic smooth DM template normalization does not translate in a proportional reduction of the IGRB. Instead, this change gets compensated by a lower IC template which, as the DM normalization increases, can go down  to almost zero in the corresponding energy bins before the IGRB measurement is altered. From this exercise, we conclude that a potential smooth Galactic DM signal could be (partially) modeled/absorbed by the IC templates, which thus may unintentionally subtract a DM signal in the IGRB measurement. This is also mostly the reason why we decided {\it not to add} any portion (\eg the high-latitude emission) of the Galactic smooth DM template to the isotropic signal, {as it was often done in literature, \eg in refs.~\cite{2010JCAP...11..041A,Calore:2013yia}}. For high-mass DM templates, however, degeneracies between the DM and the IC templates are limited, because the normalizations of the IC templates at these energies are fixed from gamma-ray data at intermediate energies (see point 2 in section~\ref{sec:reviewEGB}).

\begin{figure}[t]
\center{
\includegraphics[width=0.49\textwidth]{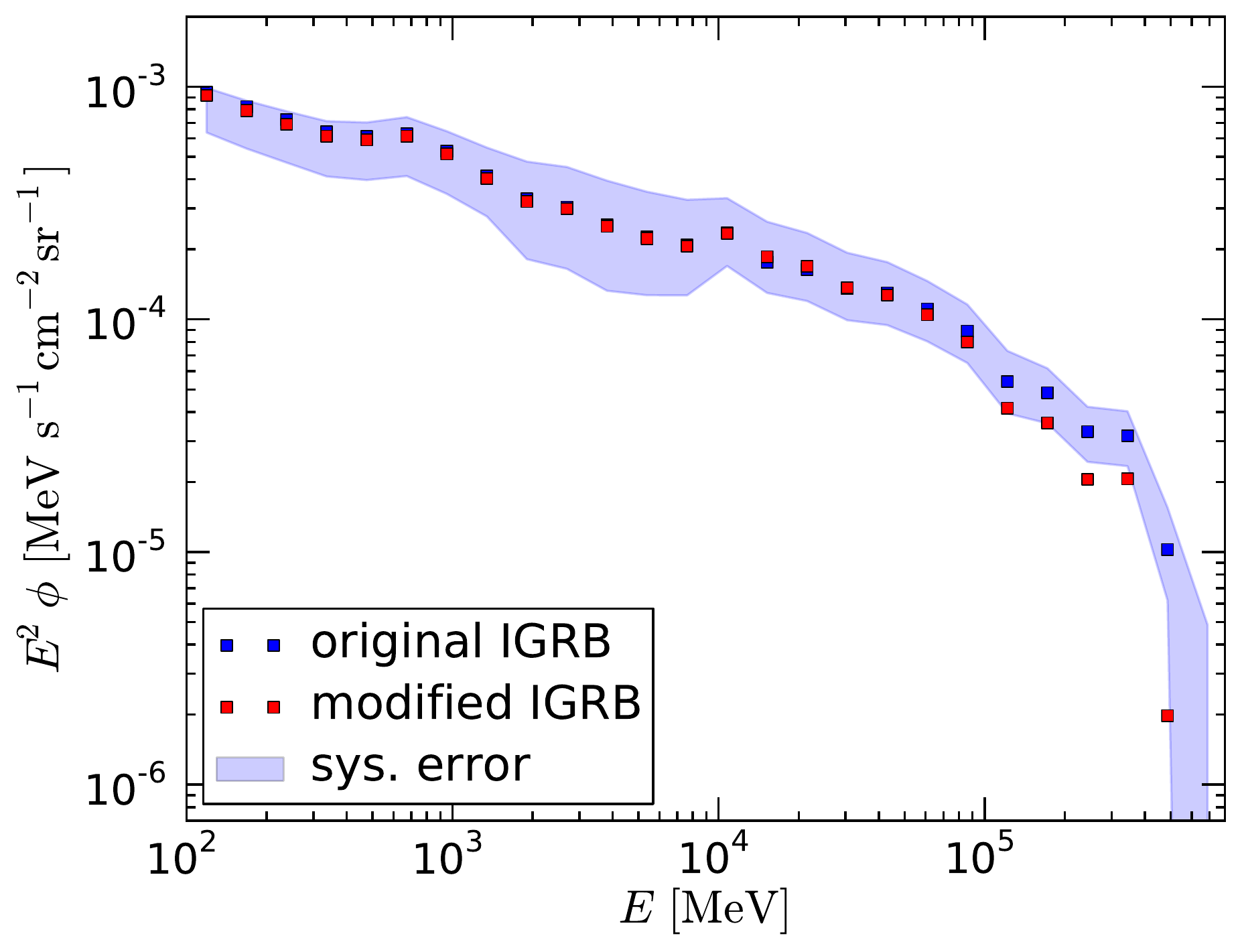}
\includegraphics[width=0.49\textwidth]{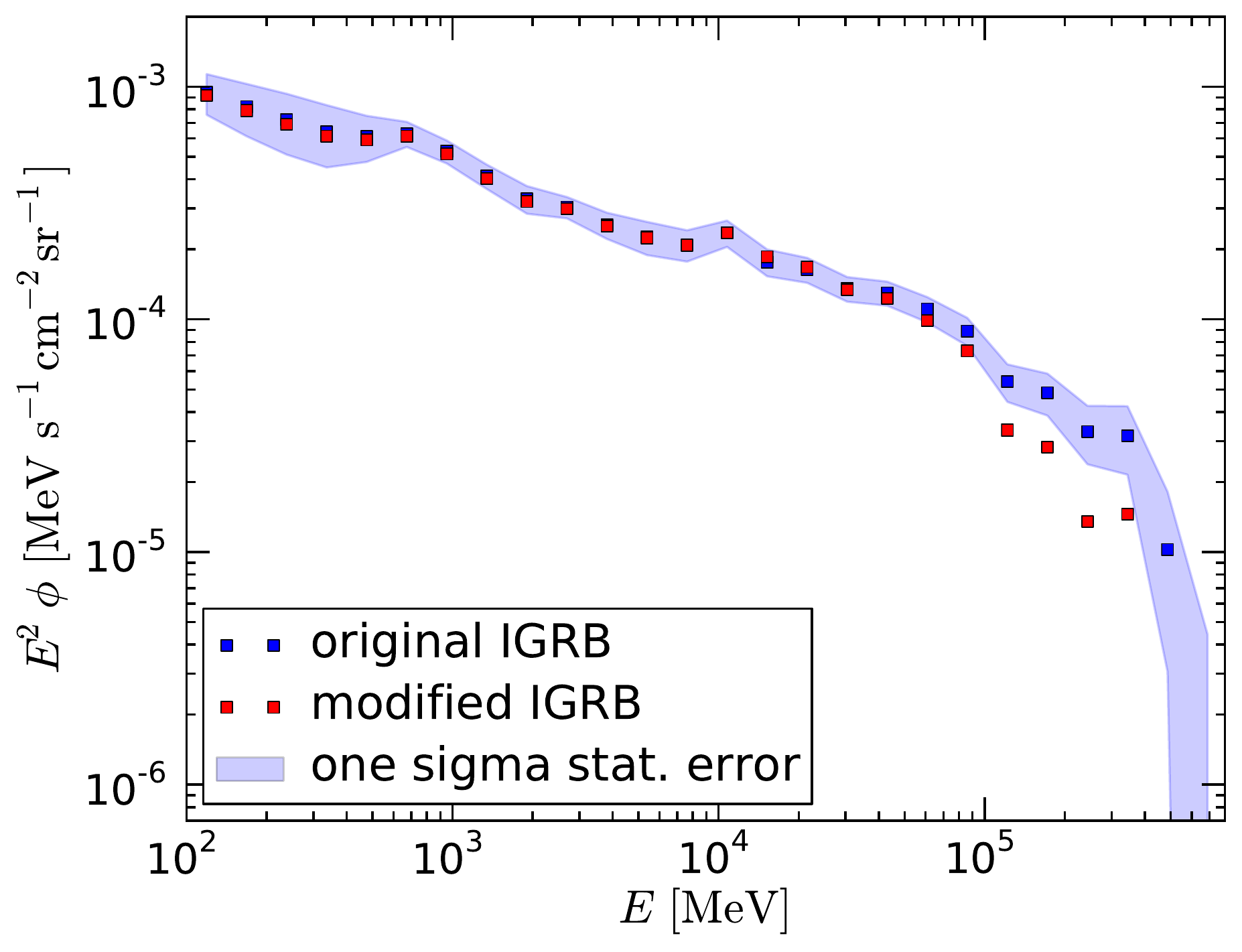}

}
\caption{{
{\em Left:} 
The new IGRB measurement, after the inclusion of the Galactic smooth DM template, when the measurement for some energy bins falls outside {twice} 
the \emph{systematic uncertainty} band, defined as the scatter among the different IGRB spectra derived in ref.~\cite{EGBnew} (the case to be compared with our {\it conservative} limits). {\em Right:}
The modified IGRB, after the inclusion of the DM template, when the measurement for some energy bins falls outside {\emph{two times} 
the \emph{$1\sigma$} statistical error band} of the IGRB measurement originally presented in ref.~\cite{EGBnew} {(to be compared with our calculation of the {\it sensitivity reach})}. {A {5 TeV  WIMP} which annihilates promptly into $b{\bar b}$ was used in both panels, which also explains why the maximum differences between the original and modified IGRB are found around {200 GeV} in these particular examples.} {Note that we do not show the statistical error bar of the modified IGRB because it is not relevant for our determination of the modified IGRB.}
}}
\label{fig:new_IGRB}
\end{figure}

Note that this prescription does not guarantee that  morphological residuals are kept small, since the methodology only investigates deviations in the IGRB spectrum. By also using the morphological information we could impose tighter constraints on a Galactic DM annihilation signal because large-scale anisotropic residuals could potentially start to appear in the IGRB measurement. In figure~\ref{fig:new_IGRB}, {we show two examples of a changed IGRB spectrum for the two cases mentioned above in the case of a 1 TeV WIMP annihilating into $b{\bar b}$.} 
%
%%%%%%%%%%%%%%%%%%%%%%%%%%%%%%%%%%%%%%%%%%%%%%%%%%%%%%%%%%%%%%%%%%%%%%%%
%%%%%%%%%%%%%%%%%%%%%%%%%%%%%%%%%%%%%%%%%%%%%%%%%%%%%%%%%%%%%%%%%%%%%%%%
%
%
%%%%%%%%%%%%%%%%%%%%%%%%%%%%%%%%%%%%%%%%%%%%%%%%%%%%%%%%%%%%%%%%%%%%%%%%
\begin{figure}[ht!]
\center\includegraphics[width=0.99\textwidth]{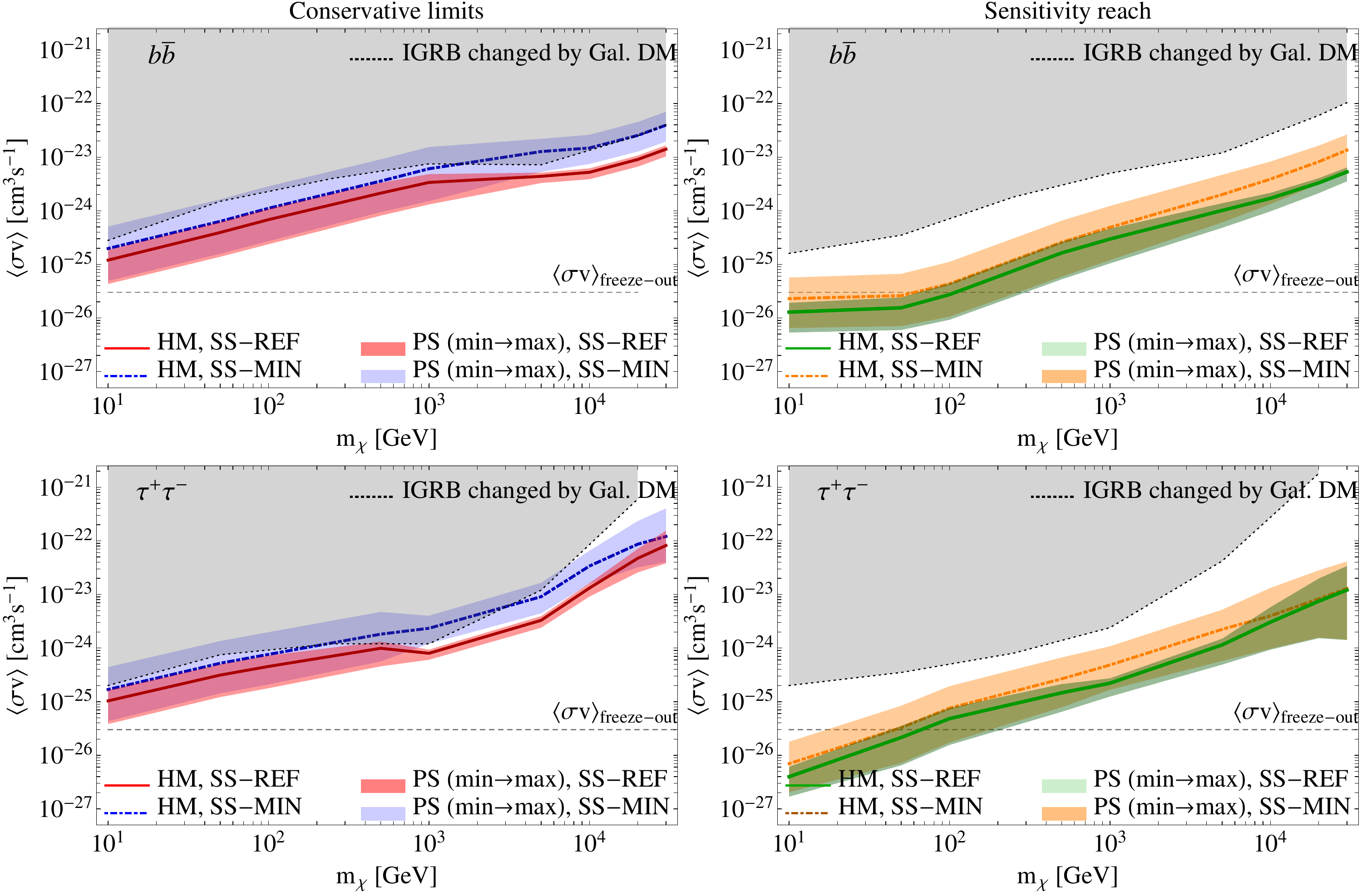}
\caption{
{{The gray regions above the dotted lines indicate the DM annihilations cross sections which would alter the measured IGRB spectra significantly due to the signal from smooth DM halo component of the Milky Way; see section~\ref{sec:gadgetchecks}. Top and bottom panels are for $b{\bar b}$ and $\tau ^+ \tau ^-$ channels, respectively. The DM limits shown are the same as those presented in figure~\ref{fig:CONSlimits_sigv} (left panels) and figure~\ref{fig:OPTlimits_sigv} (right panels).}}}\label{fig:bad_BF_norms}
\end{figure}

Figure~\ref{fig:bad_BF_norms} shows the largest possible DM annihilation cross sections to the $b{\bar b}$ and $\tau ^+\tau^-$ channels which \emph{do not} change the IGRB spectrum, together with our conservative limits on the cross section and sensitivity reach derived in section~\ref{sec:limits}.\footnote{Only model A is used in the figure, but we note that similar results are obtained with models B and C.} {The {\it non-gray-shaded} areas in figure~\ref{fig:bad_BF_norms} roughly indicate the regions where our method of deriving limits on an isotropic DM signal would not lead to significantly altered results due to the modified IGRB measurement from the presence of the assumed smooth Galactic DM signal.}

{Notably, there are regions of the parameter space where DM limits overlap with the shaded areas of our \emph{conservative} limits in figure~\ref{fig:bad_BF_norms}. Inclusion of the Galactic smooth DM template can lead to both smaller and larger IGRB intensities around the DM signal peak than the one reported in ref.~\cite{EGBnew}.} 
{For some DM masses $\lsi$ 250 GeV the IGRB can \eg get {\it higher} by up to $\sim 40$\% after the inclusion of the DM template, which would naively  weaken the limits by roughly this amount. For larger DM mass ($\gsi$ 1 TeV) the IGRB spectrum is typically lowered. This is a consequence of our procedure in which the normalizations of the Galactic foreground spectra are determined at  energies lower than the DM signal peak and then kept fixed at the higher energies. The measured IGRB intensity, the normalization of which is free in all energy bins, is therefore lowered around the energies where the  DM signal peaks in order to accommodate the presence of the Galactic DM signal.} 

{
Cross sections at the level of the sensitivity reach of the IGRB measurement are found to lie below their gray shaded region. We check the impact of using modified IGRB models derived under the assumptions that a Galactic DM signal is present. These alternate IGRB models are derived as above, with the Galactic DM signal fixed by the annihilation channel and cross section (the DM density profile is kept to the same as before). We adopt the cross-section values at the upper edge of the orange band in the top right panel of figure~\ref{fig:bad_BF_norms} (the `PS(min), SS-MIN' case) and then apply our procedure to find the sensitivity reach: we find that the cross-section sensitivity curve is basically unchanged by the inclusion of the Galactic DM component.} {For cross sections within the gray shaded area the IGRB is sometimes no longer described well by the adopted background model, so the method is no longer expected to behave well.}

Note that while our DM limits depend on the substructure signal strength {and the assumed minimal DM halo mass}, the shaded gray region in figure~\ref{fig:bad_BF_norms} is independent of it, {so the relative position of the gray region and the limits would be different for a different choice of these parameters}.

In order to exhaustively explore the impact of  Galactic smooth DM templates on the derivation of the IGRB, a larger number of Galactic astrophysical emission models should be studied. In this way it would be possible to probe in detail the IGRB along with various Galactic DM signals. However, such studies are beyond the scope of this work, which is tied to the methodology used in ref.~\cite{EGBnew}.  The initial study performed in this section shows the importance of including the Galactic DM annihilation with its proper morphology in a detailed study of isotropic intensities.\footnote{Note that the morphology of a DM template is relevant when performing the full-sky fits, since it influences the normalization of the isotropic template, \ie, the IGRB spectrum. This should not be confused, however, with the fact that we define the normalization of the DM template by the requirement that solely the IGRB {\it spectrum} gets changed, and not when the {\it whole-sky residuals} worsen significantly.} Note that this issue is more severe for decaying DM models {(studied in this context in e.g. \cite{Cirelli:2009dv,Cirelli:2012ut})}, since the DM Galactic component is more isotropic compared to the annihilating DM case.\footnote{More precisely, the smooth Galactic DM signal varies by factors of $\sim 16$ and  $\sim 4$ for annihilation and decay, respectively, between Galactic latitudes 20\de and 90\de.} {Also, the Galactic DM signal at high latitudes is typically larger than the corresponding cosmological one \cite{Ibarra:2013cra}, when compared to the annihilating DM case (see figure~\ref{fig:CONSlimits_fluxes}), and therefore a careful study of the degeneracy between the DM and IC templates becomes mandatory before robust upper limits can be determined for any potential cosmological decaying DM signal. }

%% file: 5Summary.tex
\section{Summary} \label{sec:summary}

We utilize the recent measurement of the IGRB spectrum, based on 50 months of {\fermi~\lat} data, to set limits on the isotropic DM annihilation signals, \ie the gamma rays originating from DM annihilation in halos over all of cosmic history as well as from the Galactic subhalos. Thanks to the broad energy range covered by the new IGRB measurement presented in ref.~\cite{EGBnew}, which extends up to 820 GeV, we are able to effectively constrain signals from annihilation of DM particles with masses ranging between a few GeV and a few tens of TeV. 

{At the lowest WIMP masses, our conservative DM limits in figure~\ref{fig:CONSlimits_sigv} reach thermal cross-section values for $b {\bar b}$ and $\tau^+ \tau^-$ channels.  For the case of our benchmark values for both the DM cosmological and Galactic substructure signals, the sensitivity reach of the IGRB measurement shown in figure~\ref{fig:OPTlimits_sigv}  
is comparable to the limits recently obtained using {\lat} observations of dwarf spheroidal galaxies \cite{2015arXiv150302641F,Geringer-Sameth:2014qqa} as well as {those derived from low Galactic latitude data \cite{Ackermann:2012rg}.}}

{For {WIMP} masses above 5 TeV, our conservative limits calculated for the benchmark values of the DM signal, are a factor of a few better than the ones presented in the {\fermi~\lat Collaboration} works cited above.  At these high {WIMP} masses ($\gsi$ 1 TeV), the benchmark conservative limits are comparable to those obtained from observations of dwarf spheroidal galaxies by ground-based gamma-ray telescopes (more precisely, the recent observation of Segue~1 by both \veritas \cite{Aliu:2012ga} and \magic \cite{Aleksic:2013xea}), but weaker than the limits derived from the Galactic center halo by \hess \cite{Abramowski:2011hc}. The potential sensitivity to DM annihilation signals with the current IGRB measurement might reach an order of magnitude lower cross sections in this same WIMP mass range, in the case of optimal knowledge of some still uncertain (non DM) astrophysical factors. }

{Our derived predictions for the strength of an isotropic DM annihilation signal} can be considered {realistic and not over estimated}: the extrapolations performed below the resolution of current N-body cosmological simulations---necessary to account for the smallest halos---have been done in a {physical} and theoretically well-motivated way, and uncertainties {in the expected} DM signal were estimated using a well suited and complementary approach based on the non-linear matter power spectrum which is measured in N-body simulations. Furthermore, for the first time we have quantified how the IGRB measurement is affected by a Galactic foreground DM signal and thus when the latter starts to impact the derived constraints on an isotropic DM signal.

{When compared to the {earlier \fermi~\lat Collaboration work \cite{Abdo:2010dk}, which derived DM limits from the first-year IGRB measurement, the conservative limits are now about a factor of two stronger in the WIMP mass range 1~GeV to 1~TeV for the {\it same value} of the flux multiplier $\zeta(z)$.} This improvement can be attributed to the new IGRB data used and, most notably, to the fact that we did not take into account the (isotropic) signal from the Galactic substructure in the previous work.}

{Moreover, the uncertainties of the flux multiplier $\zeta(z)$ have considerably shrunk in the present work, and it now has a factor $\sim 20$ uncertainty when the minimal halo mass cut-off is set to $10^{-6}$  h$^{-1}~\Msun$.} We note that this theoretical uncertainty range is a factor $\sim 5$ smaller than in ref.~\cite{Abdo:2010dk}. We did not consider extreme power-law extrapolations of the many relevant quantities in the Halo Model framework, which previously resulted in an over estimation of the predicted strength of the isotropic DM signal. {In this work, the theoretical uncertainty range for the predicted DM signal strength (for a given WIMP annihilation cross section) therefore covers only the lower and more physically motivated part of the previously considered range in ref.~\cite{Abdo:2010dk}.}
This in turn implies that our limits are generally consistent with the most conservative estimates derived in other works \cite{Abdo:2010dk,Cholis:2013ena,Calore:2013yia}.

In our work, we identified and addressed three main sources of uncertainty affecting the {derived limits to the DM annihilation cross section,} which can be summarized as follows: i) theoretical predictions for the strength of the DM annihilation signal, which stem on one hand from the modeling of DM clustering at small scales, and which translate into an uncertainty of a {factor $\sim 20$} (see section~\ref{sec:HM&PS}) and, on the other, from the precise amount of substructure in the Galaxy, which has a factor $\sim 3$ uncertainty, ii) modeling of the contribution of unresolved extragalactic sources to the IGRB, {reflected in the difference between our conservative limits and our derived sensitivity reach, which for $b{\bar b}$  and $\tau^+ \tau^-$ channels and our reference prediction of the isotropic DM annihilation signal ranges by a factor $\sim 3-26$, depending on the WIMP mass range considered, and iii) modeling of the Galactic diffuse emission, which can lead to variations in the limits by a factor of up to $\sim 3$.

We also studied the impact on the IGRB measurement of a DM signal from the Milky Way DM halo, and defined a region in the cross section versus DM mass plane for which the IGRB measurement would be sufficiently changed by the presence of Galactic DM as to potentially  affect the limits derived here.

{With these considerable uncertainties in mind, at present the IGRB does not represent a clean target to {\it search} for a DM signal. At the same time, we showed in figure~\ref{fig:OPTlimits_sigv} that the method has the great potential to test the `vanilla' WIMP paradigm (\ie the thermal cross-section value) up to masses of a few tens of GeV, making this approach competitive with other DM probes.} An additional strength lies in offering a complementary and truly cosmological probe for any potential DM signal hint that might be claimed from another indirect, direct or collider search.

In the coming years of the \fermi mission, the \lat sensitivity to point-like sources will improve due to the increased exposure and improved event classification and reconstruction, which, depending on the energy band, will translate into a lower IGRB intensity and better DM limits. This can be the case at energies above 100 GeV, where the total IGRB might ultimately be attributable to \fermi~\lat-detected point-like sources (as suggested in ref.~\cite{EGBnew}, where the cumulative intensity from 2FGL sources and the IGRB are seen to be comparable at the highest energies). On the other hand, at energies below 100~GeV, resolving more point sources in the next several years is not expected to significantly lower the IGRB flux (since, \eg, {new} blazars will be extremely faint and thus their contribution will be hard to detect {by the \lat,} see \cite{2010ApJ...720..435A}). 
Most importantly, developments are foreseen to help in lowering the most critical uncertainties:
i) the Euclid satellite \cite{2011arXiv1110.3193L} and next generation N-body cosmological simulations (\eg \cite{Guedes:2011ux,2012arXiv1206.2838A}), by shedding light on the small-scale DM clustering properties, ii) new constraints on the contribution of astrophysical sources to IGRB, from anisotropy and cross correlation studies of high latitude gamma-ray emission, and iii) more precise cosmic-ray measurements (with \eg AMS-02 \cite{Accardo:2014lma}) as well as detailed Planck dust maps, that will help in refining the modeling of the Galactic foreground.

%% file: appendixA_limitsABC.tex
\section{Diffuse foreground models and their impact on limits}  \label{appendix:limitsB}

In this Appendix, we investigate how the limits on the DM annihilation cross section depend on the particular foreground model that is used to derive the IGRB in ref.~\cite{EGBnew}. In section~\ref{sec:wimpconstraints} the baseline foreground emission model~A in ref.~\cite{EGBnew} was assumed for computing the DM limits. However, two more foreground models, B and C, were also defined and considered in \cite{EGBnew}, which led to slightly different derived spectra of the IGRB. These three reference models differ in propagation and CR injection scenarios, and have been derived from a customized version of \galprop. Model A is the basic reference model, whereas model B includes, \eg, an additional population of electron-only sources near the Galactic center and model C allows the CR diffusion and re-acceleration to vary significantly throughout the Galaxy. Furthermore, variations of model A have been studied, by, \eg,  changing the size of CR halo between 4 kpc and 10 kpc, modifying its CR-source distribution, turning off re-acceleration and adding {a `Fermi Bubbles'} template to assess the systematic uncertainties in the derived IGRB.

\begin{figure}[htbp]
\centering
\includegraphics[width=0.99\textwidth]{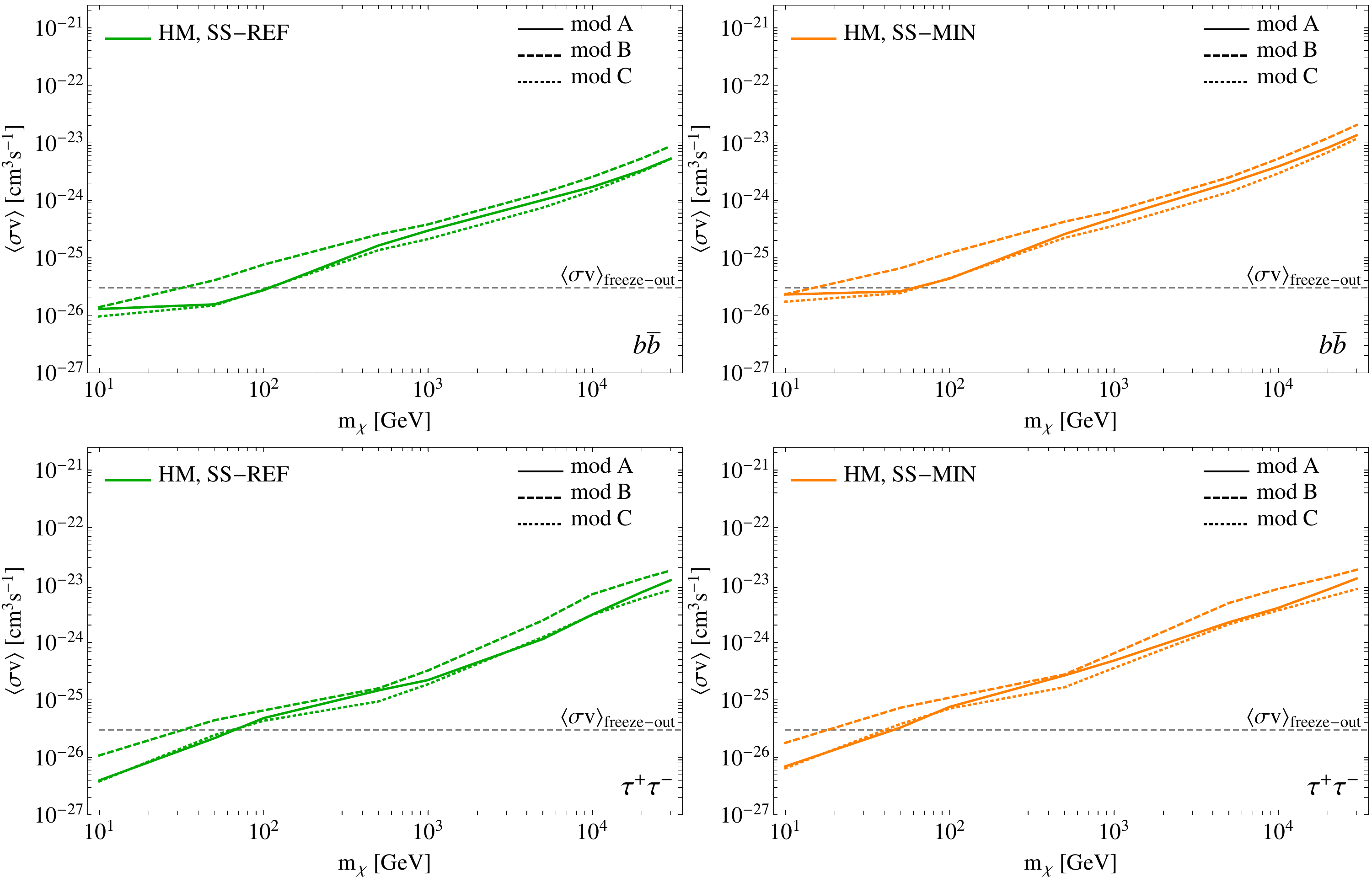}
\caption{{DM limits for the three IGRB measurements derived assuming foreground emission models A, B and C in ref.~\cite{EGBnew}. The limits shown in section~\ref{sec:wimpconstraints} implicitly assume foreground model A in ref.~\cite{EGBnew}. The limits are substantially modified when assuming model B instead, especially at low masses. This figure is for the particular case of the sensitivity reach procedure (as it was described in section~\ref{sec:ConOpt_approach}) for DM particles annihilating into $b{\bar b}$ quarks ({\it top}) and $\tau ^+ \tau ^-$ ({\it bottom}), and for the two scenarios of the Galactic substructure contribution introduced in section~\ref{sec:GalacticDM}. The case of our reference substructure model is shown on the {\it left} panels, while the minimal substructure case is on the {\it right}.}}
\label{fig:limitsB}
\end{figure}

In figure~\ref{fig:limitsB}, we show the  effect on the limits when assuming {models B and C} for the foreground diffuse emission instead of model A, {for the particular case of our sensitivity reach estimate and the $b{\bar b}$ and $\tau^+\tau^-$ annihilation channels. }

The limits are substantially modified for low mass WIMPs, \ie, below $\sim 300$ GeV, although the limits are again rather comparable at the lowest DM masses considered, namely below $\sim 20$ GeV. The maximum difference between the limits is found at about 50 GeV, where model A makes the limits a factor $\sim 2.7$ more stringent than the ones deduced using model B (assuming our fiducial Galactic substructure scenario). In all cases, these differences can be explained as the interplay between the measured IGRB  points obtained in ref.~\cite{EGBnew} for each foreground model and the WIMP mass considered. For example, in the particular case of a 50 GeV WIMP, for which the emission peaks at roughly a few GeV, there is a downward {trend} of the IGRB data between 2 GeV and 10 GeV in models A and C, which is not present in model B, making the limits substantially stronger for models A and C compared to B.

It is worth emphasizing that, in contrast, the conservative limits {`by construction'} take into account the variation induced on the IGRB measurement from using the different foreground models. In this case, we recall that, in order to set the limits, we shifted the IGRB data points to the maximum allowed intensity values among those given by the various foreground models, which always results in the most conservative bounds.

%
%%%%%%%%%%%%%%%%%%%%%%%%%%%%%%%%%%%%%%%%%%%%%%%%%%%%%%%%%%%%%%%%%%%%%%%%
%%%%%%%%%%%%%%%%%%%%%%%%%%%%%%%%%%%%%%%%%%%%%%%%%%%%%%%%%%%%%%%%%%%%%%%%
%

%% file: appendixB_3sigma.tex
\section{Limits at different confidence levels}  \label{appendix:3sigma}
\begin{figure}[htbp]
\centering
\includegraphics[width=0.93\textwidth]{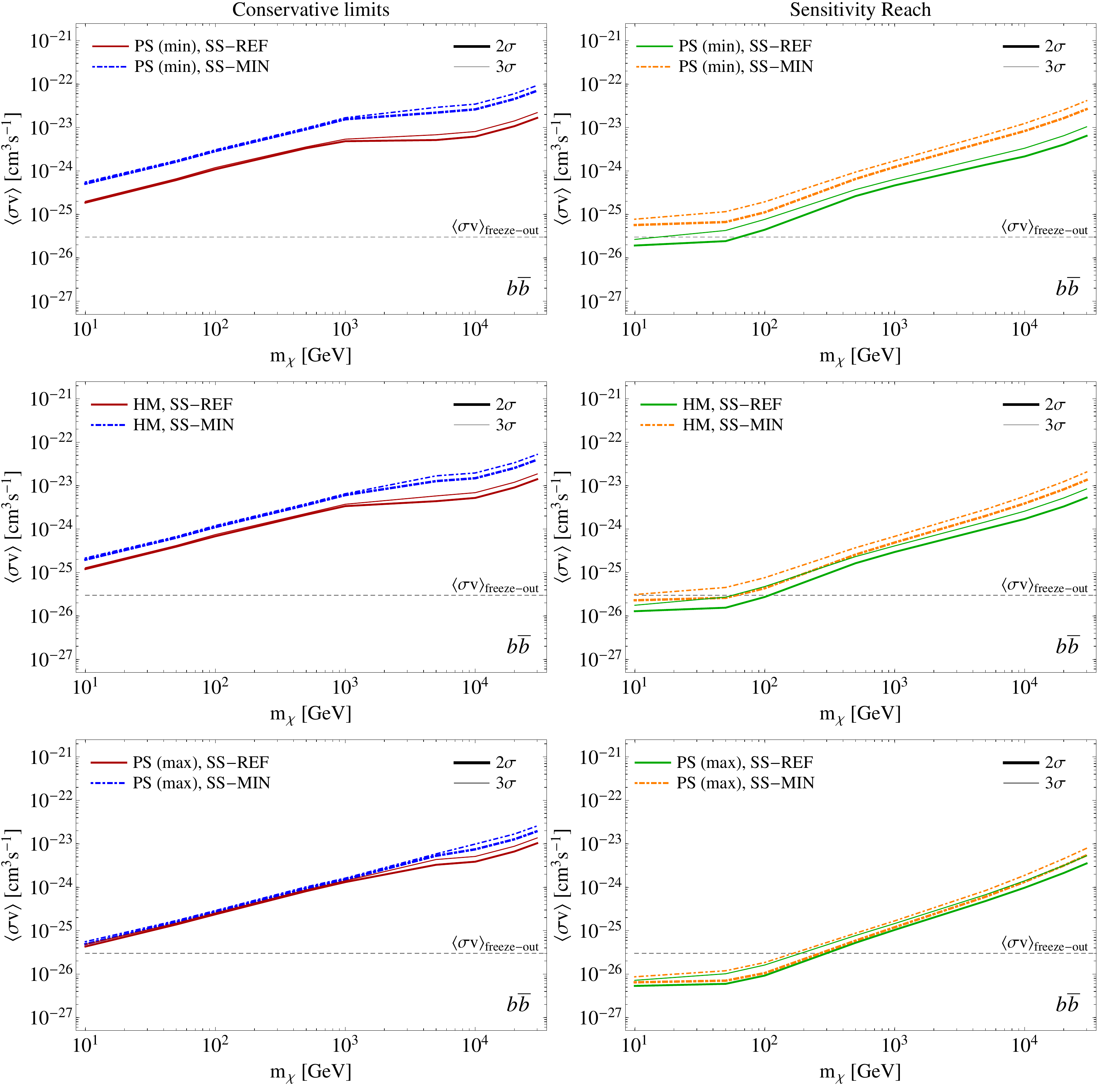}
\caption{{Upper limits on the self-annihilation cross section obtained in our conservative ({\it left}) and sensitivity reach ({\it right}) procedure. The limits are on the annihilation cross section into $b\bar{b}$ quarks and the thicker (thinner) lines show the 2$\sigma$ (3$\sigma$) limits. The figures in each row show the limits in different DM setups: (top) the minimal extragalactic signal in the PS approach, (middle) the benchmark extragalactic signal in the HM approach, and (bottom) the maximal extragalactic signal in the PS approach. In each figure, the solid line corresponds to the benchmark Galactic substructure intensity, while the dashed line represents the minimal assumed Galactic substructure signal; see section~\ref{sec:GalacticDM}. The minimal scale for DM structures corresponding always to a halo mass cut-off of  $\Mmin=10^{-6}$  h$^{-1}\Msun$.}}
\label{fig:3sigma}
\end{figure}

\begin{figure}[htbp]
\centering
\includegraphics[width=0.99\textwidth]{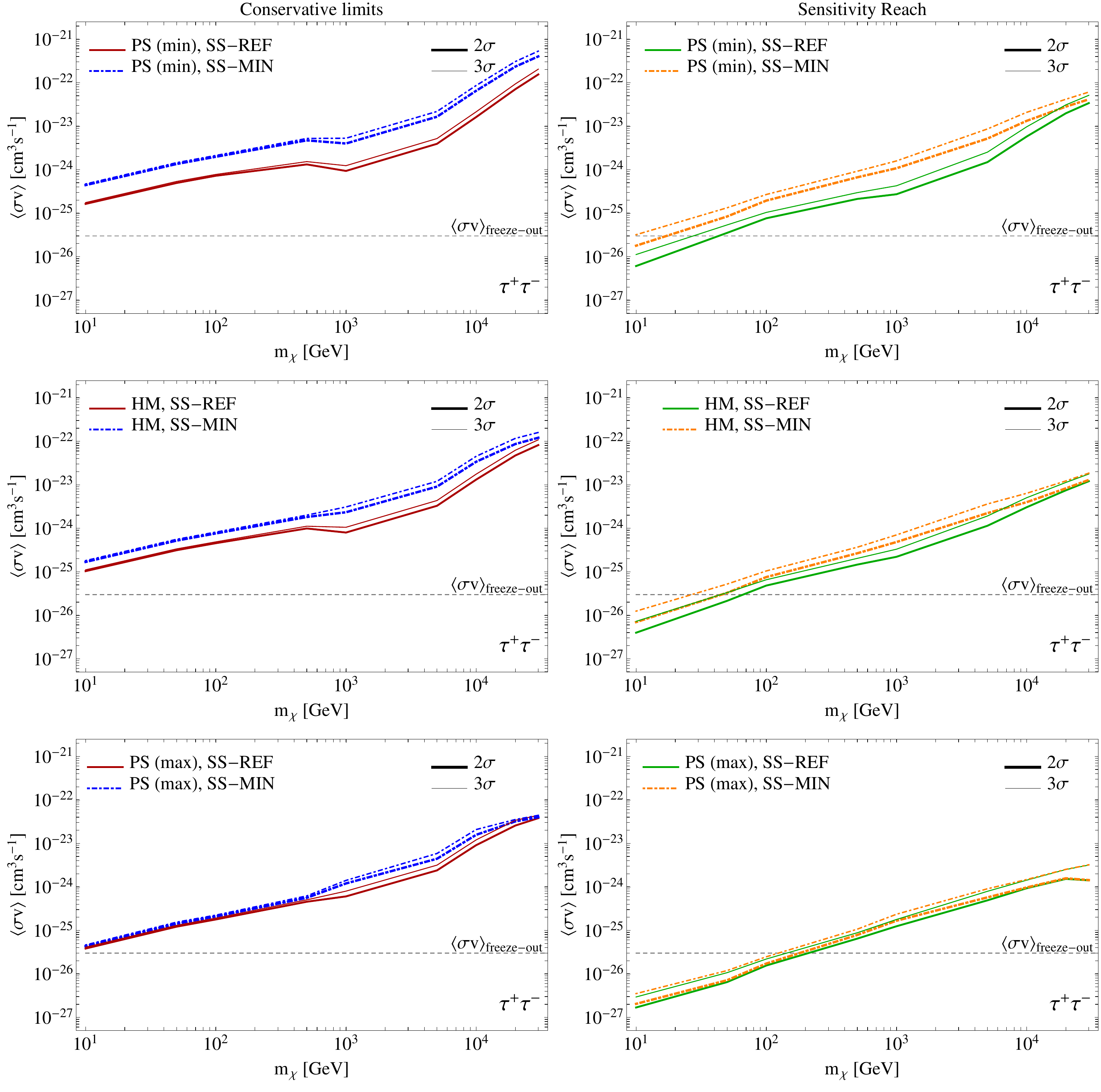}
\caption{{Same as figure~\ref{fig:3sigma}, but for the $\tau^+\tau^-$ channel.}}
\label{fig:3sigmatau}
\end{figure}

In this Appendix we compare 2$\sigma$ and 3$\sigma$ upper limits on the DM annihilation cross section for $b {\bar b}$ and $\tau^+ \tau^-$ channels (figures~\ref{fig:3sigma} and \ref{fig:3sigmatau}, respectively), for both {the conservative limits and the sensitivity reach,} and for the six representative cases of the DM signal strength considered in our work.
%
%%%%%%%%%%%%%%%%%%%%%%%%%%%%%%%%%%%%%%%%%%%%%%%%%%%%%%%%%%%%%%%%%%%%%%%%
%%%%%%%%%%%%%%%%%%%%%%%%%%%%%%%%%%%%%%%%%%%%%%%%%%%%%%%%%%%%%%%%%%%%%%%%
%
%
%%%%%%%%%%%%%%%%%%%%%%%%%%%%%%%%%%%%%%%%%%%%%%%%%%%%%%%%%%%%%%%%%%%%%%%%
%%%%%%%%%%%%%%%%%%%%%%%%%%%%%%%%%%%%%%%%%%%%%%%%%%%%%%%%%%%%%%%%%%%%%%%%
%